\documentclass[aps,prx,onecolumn,superscriptaddress,footnote,notitlepage]{revtex4-1}

\usepackage{relsize}
\usepackage{graphicx}
\usepackage{bm}
\usepackage{amsmath, bbm}
\usepackage{amssymb}
\usepackage{mathrsfs}
\usepackage{xcolor}
\usepackage{cancel}
\usepackage{titlesec}
\usepackage[normalem]{ulem}
\usepackage{soul}
\usepackage{dsfont}
\usepackage[caption=false]{subfig}

\titlespacing{\section}{0ex}{2ex}{0.4ex}
\titleformat{name=\section}{\bf}{\thesection.}{.3em}{}
\titlespacing{\subsection}{0ex}{2ex}{0.4ex}
\titleformat{name=\subsection}{\bf}{\thesubsection.}{.3em}{}

\newcommand{\eq}[1]{Eq.~\eqref{#1}}
\newcommand{\fig}[1]{Fig.~\ref{#1}}

\def\be{\begin{eqnarray}}
\def\ee{\end{eqnarray}}
\newcommand{\tr}{\mathrm{tr}}
\newcommand{\<}{\langle}
\renewcommand{\>}{\rangle}
\newcommand{\ket}[1]{|{#1}\rangle}
\newcommand{\bra}[1]{\langle{#1}|}
\newcommand{\pr}[1]{\Pi[{#1}]}
\newcommand{\projmap}[2]{\Pi[{#1}] \, {#2} \, \Pi[{#1}] }
\newcommand{\avg}[1]{\left\langle {#1} \right\rangle}

\newcommand{\irr}{\mathrm{irr}}
\newcommand{\tth}{{\tilde{ \theta}}}
\newcommand{\coh}{\mathit{coh}}
\newcommand{\noneq}{\mathit{nonth}}
\newcommand{\wext}{W_\mathrm{ext}}

\newcommand{\Var}[1]{\mathrm{Var}\left({#1}\right)}

\newcommand{\sub}[1]{_{\!\mathsmaller{\, #1}}}
\newcommand{\s}{{\mathcal{S}}}
\newcommand{\e}{{\mathcal{E}}}
\newcommand{\lo}{{\mathcal{L}}}
\newcommand{\vv}{{\mathcal{V}}}
\newcommand{\uu}{{\mathcal{U}}}
\newcommand{\T}{{\mathcal{T}}}
\newcommand{\bb}{{\mathcal{B}}}
\newcommand{\co}{\mathds{C}}
\newcommand{\re}{\mathds{R}}
\newcommand{\one}{\mathds{1}}

\newcommand{\iii}{{\mbox{\footnotesize (III)}}}

\newcommand{\DFiii}{\Delta F^\iii}
\newcommand{\Siii}{S^\iii_\irr}

\newcommand{\Gthr}{\Gamma^\iii}
\newcommand{\Gfour}{{\Gamma^{\mbox{\footnotesize (IV)}}}}

\newcommand{\GthreeS}{\Gamma^\iii_{(l,m, n)}}
\newcommand{\GthreeSmin}{\Gamma^\iii_{(l, n)}}

\newcommand{\GclS}{\Gamma^\mathrm{cl}_{(m,n)}}
\newcommand{\Gclwoindex}{{\Gamma^\mathrm{cl}}}

\newcommand{\Gq}{\Gamma^{\mathrm{q}}_{(l, m)}}

\newcommand{\Gqwoindex}{{\Gamma^\mathrm{q}}}

\newcommand{\sqG}{s^{\rm qu}_\irr \left(\Gq \right)}

\newcommand{\sclG}{s^{\rm cl}_\irr \left(\GclS \right)}

\newcommand{\clentwo}{\<s^{\rm cl}_\irr\>}
\newcommand{\qentwo}{\<s^{\rm qu}_\irr\>}
\newcommand{\qenttheta}{\<s^{\rm qu}_\irr (\Theta)\>}
\newcommand{\qentthetat}{\<s^{\rm qu}_\irr (\Theta, t)\>}

\newcommand{\clent}{\clentwo}
\newcommand{\qent}{\qentwo}

\newcommand{\Qq}{Q_{\rm qu}}
\newcommand{\QqG}{Q_{\rm qu}  \left(\Gq \right)}

\newcommand{\Qcl}{Q_{\rm cl}}

\newcommand{\SvN}{S_{\rm vN}}

\newcommand{\Qclavg}{\left\<\Qcl \right\>_\Gclwoindex}
\newcommand{\Qqavg}{\left\<\Qq \right\>_\Gqwoindex}
\newcommand{\Qivavg}{\left\< Q^{\mbox{\footnotesize (IV)}}_{\rm cl}\right\>_{\Gfour}}

\newcommand{\Qclavgwo}{\left\<\Qcl \right\>}
\newcommand{\Qqavgwo}{\left\<\Qq \right\>}

\newcommand{\qdiss}{Q^{\rm sur}_{\mathrm{diss}}} 
\newcommand{\wirr}{W_{\rm irr}} 

\newcommand{\DScl}{\Delta S_{\rm cl} }
\newcommand{\DSqu}{\Delta S_{\rm qu} }
\newcommand{\DSFour}{\Delta S^{\mbox{\footnotesize (IV)}}}

\newcommand{\siv}{\<s^{\mbox{\footnotesize (IV)}}_\irr\>}

\begin{document}

\title{Energetic footprints of irreversibility  in the quantum regime} 

\author{M. H. Mohammady}
\email{m.hamed.mohammady@savba.sk }
\affiliation{CEMPS, Physics and Astronomy, University of Exeter, EX4 4QL, United Kingdom.}
\affiliation{Department of Physics, Lancaster University, LA1 4YB, United Kingdom.}
\affiliation{RCQI, Institute of Physics, Slovak Academy of Sciences, D\'ubravsk\'a cesta 9, Bratislava 84511, Slovakia.}

\author{A. Auff\`eves}
\email{alexia.auffeves@neel.cnrs.fr}
\affiliation{CNRS and Universit\'e Grenoble Alpes, Institut N\'eel, F-38042 Grenoble, France.}

\author{J. Anders}
\email{janet@qipc.org}
\affiliation{CEMPS, Physics and Astronomy, University of Exeter, EX4 4QL, United Kingdom.}
\affiliation{Institute of Physics and Astronomy, University of Potsdam, 14476 Potsdam, Germany.}


\begin{abstract}

In classical thermodynamic processes the unavoidable presence of irreversibility,  quantified by the entropy production, carries two energetic footprints: the reduction of extractable work from the optimal, reversible case, and the generation of a surplus of heat that is irreversibly dissipated to the environment. Recently it has been shown that in the quantum regime an additional quantum irreversibility occurs that is linked to decoherence into the energy basis. Here we employ quantum trajectories to construct distributions for classical heat and quantum heat exchanges, and show that the heat footprint of quantum irreversibility differs markedly from the classical case. We also quantify how quantum irreversibility reduces the amount of work that can be extracted from a state with coherences. Our results show that decoherence leads to both entropic and energetic footprints which both play an important role in the optimization of controlled quantum operations at low temperature.

\end{abstract}

\maketitle

\section*{Introduction} 

In recent years much effort has been made in extending the laws of thermodynamics to the quantum regime \cite{Goold2015,Millen2016,Vinjanampathy2016,Binder2018}. Maximal work extraction (or minimal work cost) has been discussed for a range of protocols~\cite{Allahverdyan2004,Aberg2013,Frenzel2014a,Perarnau-Llobet2014, Skrzypczyk2014b, Lostaglio2015f, Cwikliski2015a,  Lostaglio2015d, Mitchison2015, Korzekwa2016b, Patia, Miller2016, Uzdin2016, Ng2016, Streltsov2016a, Frenzel2016a,Klatzow2017,Kwon2017,Mohammady2017,Morikuni2017}, showing that energetic coherences can be a resource for work extraction \cite{Uzdin2015a,Uzdin2015b, Anders-coherence-thermo, Solinas2016, Lostaglio2015b} while quantum correlations can reduce the work cost of erasing information \cite{DelRio2011}. However, many of these studies have focussed on the optimal limit of reversible processes, i.e. unitary and quasi-static evolutions, without discussing the limitations that irreversibility puts on work extraction. 
On the other hand,  the irreversibility of thermodynamic processes in the quantum regime has been explored using stochastic thermodynamics \cite{Callens2004, Horowitz-entropy-quantum-jump, Alonso2016, Francica2017a, Santos2019, Alexia-measurement-thermodynamics, Alexia-fluctuation-engineered-reservoir, Manzano2017, Manikandan2018}
leading to the notion of a fluctuating quantum entropy production \cite{Deffner2011} that obeys a fluctuation theorem analogous to those of classical non-equilibrium dynamics \cite{Crooks1999, Seifert2005,Seifert2012}. First experiments have now measured  entropy production rates in driven mesoscopic quantum systems for two platforms, a micromechanical resonator and a Bose-Einstein condensate \cite{Brunelli2018}. Most recently, the average entropy production of a quantum system that interacts with another (non-bath) system, has been shown to include an additional information flow term \cite{Ptaszynski2019}.

In classical thermodynamics irreversibility occurs whenever a non-thermal system is brought into contact with a thermal environment. The ensuing relaxation of the system leads to exchanges of energy that cannot be reversed with the same thermodynamic cost. In thermodynamics this irreversibility is quantified by the positive ``irreversible entropy production'' $S_\irr : = \Delta S - \frac{Q}{T}  \geqslant 0$,
which measures the discrepancy between the system's entropy increase $\Delta S = S_\mathrm{fin} - S_\mathrm{ini}$ during any thermodynamic process and the heat $Q$ absorbed by the system from the environment divided by the environment's temperature $T$.  Hence when a process with entropy change $\Delta S$ incurs a non-zero entropy production $S_\irr$ this results in a surplus of heat,  \cite{Landau1980}
\be \label{Qirr}
	\qdiss = T \, S_\irr,
\ee 
that is irreversibly dissipated from the system to the environment (in comparison with a reversible process resulting in the same entropy change $\Delta S$). Irreversibility also puts a fundamental bound on the amount of work $\wext $ that can be extracted during isothermal processes \cite{Landau1980,Balian2007},
\be \label{wext}
	\wext = - \Delta F - T S_\irr \leqslant - \Delta F,
\ee
where $\Delta F = F_\mathrm{fin} - F_\mathrm{ini}$ is the system's free energy increase. 
The more irreversible a process is, the less work can be extracted and the term $\wirr = T S_\mathrm{irr}$ may be called the irreversible work, or non-recoverable work \cite{Weinhold2008}. \eq{Qirr} and \eq{wext}  link  entropy production, $S_\irr$ to a surplus in heat dissipation, $\qdiss \geqslant 0$, and a reduction in work extraction, $\wext \leqslant - \Delta F$. These relationships are the well-known energetic footprints of irreversibility in classical thermodynamics.   

\medskip

 A quantum system can be out of equilibrium in two ways: by maintaining energetic probabilities that are non-thermal, and by maintaining coherences between energy levels. It has been shown that contact with the thermal environment gives rise to a classical and a quantum aspect of irreversibility \cite{Francica2017a,Santos2019}. Moreover, in addition to the exchange of energy quanta between the quantum system and the thermal environment - known as classical heat - whenever the system has ``energy coherence'' it will exhibit a uniquely quantum energy exchange known as ``quantum heat''  \cite{Alexia-measurement-thermodynamics, Alexia-fluctuation-engineered-reservoir, Alonso2016, Mohammady2017, Mohammady2019c, Buffoni2018}.  However, thus far the link between quantum entropy production and its energetic footprints has remained opaque.

In this paper we establish the energetic footprints of irreversibility in the quantum regime, arising whenever a system is brought in contact with a thermal environment. For concreteness, we here consider a specific protocol that extracts work from a quantum system's coherences in the energy basis \cite{Anders-coherence-thermo}. We first extend the protocol to capture irreversible steps that are unavoidable in any experimental implementation and which will affect heat and work exchanges. By employing the eigenstate trajectory unravelling of the open system dynamics,  where at the start and end of each dynamical process the system is assumed to be in one of the eigenstates of its time-local density matrix, we identify the distributions of classical and quantum heat, and evidence that purely quantum contributions to the entropy production are not related to the average quantum heat, in stark contrast to the classical regime, cf.~\eq{Qirr}. Instead, we show that  the average quantum entropy production, $\qent$, is linked with the variance in quantum heat,  $\Var{\Qq}$,  a quantity that has recently been connected to entanglement generation \cite{Elouard2018}. Specifically, we show that $\qent = 0$ if and only if $\Var{\Qq} = 0$, while both $\qent$ and the lower bounds to $\Var{\Qq}$ monotonically decrease under Hamiltonian-covariant channels. In the special case of qubits, this relationship becomes stronger, and we show that: (i) for the family of states $\rho$ with the same spectrum, but different eigenbases, $\qent$  and $\Var{\Qq}$ are co-monotonic with the energy coherence of the eigenbasis of $\rho$; and (ii) both  $\qent$  and $\Var{\Qq}$ monotonically decrease under the action of Hamiltonian-covariant channels that are a combination of dephasing and depolarization. Both of these strong monotonicity relationships break down for systems with a larger Hilbert space,  which we illustrate with a simple  example for a three-level system.  We also note that no such relationship exists between the average classical entropy production $\clent$ and the variance in classical heat $\Var{\Qcl}$; even in the case of qubits one does not monotonically increase with the other, and furthermore  $\clent=0$ is neither necessary nor sufficient for $\Var{\Qcl}=0$, with the latter condition only being achieved in the limit of zero temperature.  Finally, we show that the classical and quantum entropy production reduce the extractable work from coherence in equal measure, cf.~\eq{wext}. The results show that when experimental imperfections are unavoidable, any work-optimization strategy needs to consider the trade-off between a system having a certain degree of classical non-thermality or quantum coherence, or both. Besides being of fundamental importance for the development of a general quantum thermodynamics framework that includes irreversibility, these relations will also be crucial for the assessment of the energetic cost of quantum control protocols, that aim to optimize performance of computation and communication in the presence of decoherence and noise.

\section*{Results}

\subsection*{Imperfect protocol for work extraction from~coherences} \label{sec:protocol}

We here outline the protocol for optimal work extraction from coherences introduced in \cite{Anders-coherence-thermo}, and modify it so as to include imperfections  that result in both classical and quantum irreversibility. This protocol can be implemented for any $d$-dimensional system, but we shall pay special interest to the qubit case for illustrative purposes. For a $d$-dimensional quantum system with Hamiltonian $H$ and quantum state $\rho$ we denote by $(\rho, H)$ any non-equilibrium configuration of the system, and by $(\tau, H)_T$ with $\tau:= e^{-H/(k_B T)}/ Z$ and partition function $Z:= \tr[e^{-H/(k_B T)}]$ its equilibrium configuration at temperature $T$ \cite{Anders-thermo-discrete}. The protocol will involve quenching of the system Hamiltonian in $N$ discrete steps, denoted $H^{(0)} \mapsto H^{(1)} \mapsto \dots \mapsto H^{(N)}$. Moreover, $H^{(j)}$ for $j=0,1,\dots,N$ are chosen  diagonal in the same basis, i.e. only the spectrum of the Hamiltonian varies during the protocol. Specifically,  $H^{(j)} := \sum_{k=1}^d E_k^{(j)} \Pi[e_k]$,  where  $E_k^{(j)}$ are energy eigenvalues, and $\Pi[\psi] \equiv |\psi\>\<\psi|$ denotes the projection onto the pure state $\ket{\psi}$.
The system is initially prepared in an arbitrary mixed  state
\be \label{eq:angle-state}
	\rho := \sum_{l=1}^d p_l \, \Pi[\psi_l],
\ee 
with $p_l >0$ for all $l$, $\sum_l p_l =1$, and $\{\ket{\psi_l}\}$ an arbitrary orthonormal basis.  

The protocol transfers $\rho$ to the fixed final state $\eta$ chosen to have the same energetic probabilities as the initial state $\rho$ but with the energetic coherences removed \cite{Anders-coherence-thermo}, i.e.  the system's final state is
\be \label{eq:eta}
	\eta := \sum_k \projmap{e_k}{\rho} \equiv \sum_k r_k \,   \Pi[e_k], 
\ee
with $r_k := \langle e_k| \rho | e_k \rangle$ quantifying the projection of $ \rho$ onto the energy eigenstate $|e_k \rangle$. 
The optimal, reversible, implementation of the $\rho$ to $\eta$  transfer was proposed in \cite{Anders-coherence-thermo} and it was shown that the \emph{``average''} work extracted is $\avg{\wext}  =   k_B T \, (\SvN(\eta) - \SvN(\rho)) \geqslant 0$, where $\SvN$ is the Von Neumann entropy, defined as $\SvN(\rho) := - \tr[\rho \, \log \rho]$. This is in agreement with equality in \eq{wext} assuming the free energy of a quantum non-equilibrium configuration is defined as $F(\rho, H) := \tr[H \, \rho] - k_B T \, \SvN(\rho)$ \cite{Landau1980,Balian2007,Gemmer2015,Esposito2010, Manzano2018}, and realising that the state change $\rho$ to $\eta$ carries no energy change, $\Delta U =0$, and hence $\Delta F = - k_B T \, \Delta \SvN$. We remark that only the ``average'' work was provided in \cite{Anders-coherence-thermo} but no distribution of work was given with respect to which $\avg{\wext}$ is an ``average''.

Generalizing first the steps of the optimal protocol \cite{Anders-coherence-thermo} to include irreversibility will allow us to investigate the impact of entropy production on distributions of work and heat below.

The new protocol consists of the following five steps, and the state evolution is visualised for a qubit in \fig{fig:protocol}: ({\bf I}) Use a unitary $V$ to rotate the quantum system's configuration $(\rho, H^{(0)})$ into configuration $(\tilde \rho, H^{(0)})$ where $\tilde \rho := V \rho V^\dagger = \sum_l p_l \Pi[\tilde \psi_l]$. In the reversible protocol, $V$ is chosen such that $\ket{\tilde \psi_l} := V \ket{\psi_l}$ is a Hamiltonian eigenstate, i.e. $[\tilde \rho, H^{(0)}] =0$  \cite{Anders-coherence-thermo}. Here we allow $V$ to be imperfect and hence $[\tilde \rho, H^{(0)}]  \ne 0$; ({\bf II}) Change the Hamiltonian rapidly resulting in a quench from $(\tilde \rho, H^{(0)})$ to $(\tilde \rho, H^{(1)})$. In the reversible protocol, the energetic levels of $H^{(1)}$ are chosen such that the configuration $(\tilde \rho, H^{(1)})$ is thermal at temperature $T$  \cite{Anders-coherence-thermo}. This is possible because we assume that we can perform arbitrary quenches of the Hamiltonian, and since the initial state $\rho$ has full rank, there exists some Hamiltonian with respect to which an energy incoherent state $\tilde \rho$ will be thermal. Here we consider the case that the energetic levels of $H^{(1)}$ are adjusted imperfectly, and hence configuration $(\tilde \rho, H^{(1)})$ is not necessarily thermal even if $[\tilde \rho, H^{(1)}]=0$; ({\bf III}) Put the quantum system in thermal contact with a heat bath at temperature $T$, 
and wait for a sufficiently long time so that $(\tilde \rho, H^{(1)})$ is brought into the thermal configuration $(\tau_1, H^{(1)})_T$; ({\bf IV}) Change the system's Hamiltonian slowly from $H^{(1)}$ to $H^{(N)}$, keeping the system in thermal contact with the heat bath. The evolution is chosen quasi-static (i.e. very slow), such that thermal equilibrium at $T$ is maintained throughout this step. The final Hamiltonian $H^{(N)}$ is chosen so that the system's thermal state is the desired final state, i.e.,  $\tau_N = \eta$; ({\bf V}) Decouple the system from the thermal bath and quench the Hamiltonian back to $H^{(0)}$, changing the system's configuration from $(\eta, H^{(N)})_T$ to the desired configuration $(\eta, H^{(0)})$.

Since Steps {\bf (I), (II), (IV)} and {\bf (V)} are either unitary or quasi-static, they are thermodynamically reversible. The thermodynamic irreversibility of the protocol occurs when the quantum system is put in contact with the thermal bath in Step  {\bf (III)}. The irreversible thermalization $(\tilde \rho, H^{(1)}) \to (\tau_1, H^{(1)})_T$ leads to a reduction in free energy, i.e. $\DFiii = - k_B T \, D[\tilde \rho\| \tau_1]$ where $D[\tilde \rho \| \tau_1] =  \tr[\tilde \rho \left(\log \tilde \rho -  \log \tau_1 \right)] \geqslant 0$ is the  quantum relative entropy between the state before thermalization, $\tilde \rho$, and the state after thermalization, $\tau_1$, which vanishes if and only if $\tilde \rho = \tau_1$. Observing that no work is exchanged during  thermalization ($\wext = 0$), and based on the assumption that \eq{wext} holds in the quantum regime, the term $k_B D[\tilde \rho \| \tau_1]$ is often identified with the  entropy $\Siii$ that is produced during the thermalization step \cite{Deffner2010, Santos2019}.

As recently discussed in \cite{Francica2017a,Santos2019}, the geometric measure of irreversibility given by the relative entropy splits into a quantum and a classical part, 
\be \label{eq:avgrelE}
	D[\tilde \rho\| \tau_1] = D[\tilde \rho\| \tilde \eta] + D[\tilde \eta \| \tau_1],
\ee 
where in analogy with \eq{eq:eta}, we define $\tilde \eta := \sum_k \projmap{e_k}{\tilde \rho}$. As we will show below, \eq{eq:avgrelE} can be obtained as averages over the entropy produced along decoherence trajectories and classical thermalization trajectories \cite{Santos2019}.
This splitting reflects the fact that the quantum configuration $(\tilde \rho, H^{(1)})$ is out of equilibrium in two distinct ways: it can have  quantum coherences between energy levels,  and classical non-thermality due to non-Boltzmann probabilities for the energies. In particular, $D[\tilde \rho\| \tilde \eta] \equiv \SvN(\tilde \eta) - \SvN(\tilde \rho)$ is known in the literature as the ``relative entropy of coherence'' which quantifies the coherence (or asymmetry) of the state $\tilde \rho$ with respect to the Hamiltonian $H^{(1)}$ \cite{Baumgratz2014, Marvian2014}. Similarly, $D[\tilde \eta \| \tau_1]$ can be seen as a measure of classical non-thermality.

\subsection*{A special case: qubits}

\begin{figure}[t]
\includegraphics[trim=0cm 0cm 0cm 0cm,clip, width=0.32\textwidth]{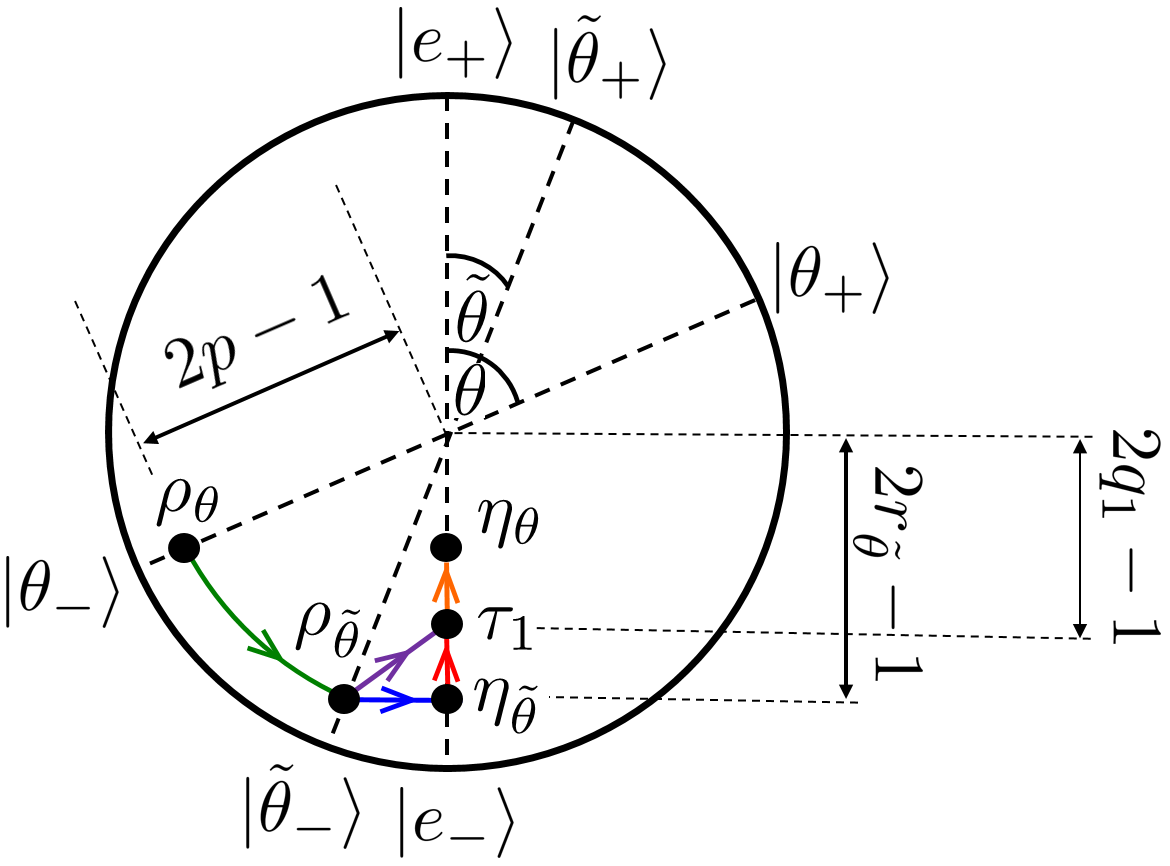}
\caption{\sf 
{\bf State evolution during work extraction protocol for qubits.} 
Initially, the system is prepared in state $\rho_\theta$ and is then unitarily evolved to $\rho_\tth$  (green arrow) which has the eigenstates $\ket{\tth_\pm}$.  Following a Hamiltonian quench  that changes the splitting of the energetic levels but does not alter the energy eigenstates $\ket{e_\pm}$, the system is put in thermal contact with a bath and allowed to relax to the thermal state $\tau_1$.  The full thermalization step (purple arrow) can be split into quantum decoherence with respect to the energy eigenbasis  (blue arrow)  $\rho_\tth \mapsto \eta_\tth$ followed by classical thermalization (red arrow) $\eta_\tth \mapsto \tau_1$. Next, the state transfer $\tau_1 \mapsto \eta_\theta$  (orange arrow) is effected by a quasistatic isothermal process. Finally, the Hamiltonian is quenched back to its initial configuration. This protocol realises the thermodynamic removal of coherences, i.e. transforming $\rho_\theta$ to $\eta_\theta$, while irreversibility arises due to the mismatches between $\rho_{\tth}$ and $\eta_{\tth}$ as well as $\eta_{\tth}$ and $\tau_1$.
\label{fig:protocol} }
\end{figure}

For the special case of qubits, we may provide an intuitive illustration of the protocol in a geometric fashion by use of the Bloch sphere. Specifically, we shall denote the $j$\textsuperscript{th} Hamiltonians as $H^{(j)} := \frac{1}{2}\hbar \omega_j(\Pi[e_+] - \Pi[e_-])$, and  represent the initial and unitarily evolved states, $\rho$ and $\tilde \rho$, in terms of angles $\theta$ and $\tth$, respectively:
\begin{align}
\rho \equiv \rho_\theta &:= p \Pi[\theta_-] + (1-p) \Pi[\theta_+], &\tilde \rho \equiv \rho_\tth := p \Pi[\tth_-] + (1-p) \Pi[\tth_+], 
\end{align}
where $1 > p > \frac{1}{2}$, and
\begin{align}
\label{eq:angular-states}
\ket{\theta_\pm} &:= \cos(\theta/2)\ket{e_\pm} \pm e^{\pm i \phi} \sin(\theta/2) \ket{e_\mp}, &\ket{\tth_\pm} := \cos(\tth/2)\ket{e_\pm} \pm e^{\pm i \tilde\phi} \sin(\tth/2) \ket{e_\mp}.
\end{align}
We note that, without loss of generality, we may assume that $\phi= \tilde\phi=0$ due to the invariance of the work extraction protocol with respect to unitary evolution generated by $H$, while $\theta$ and $\tth$ may be assumed to   fall in the range $[-\pi/2, \pi/2]$, since angles outside this range would be accounted for by changing the sign of the Hamiltonian. The decohered state $\eta$ is thus defined as $\eta_\theta = r_\theta \Pi[e_-] + (1-r_\theta) \Pi[e_+]$  with $r_\theta := \<e_-| \rho_\theta |e_-\>$, and $\eta_\tth$ is similarly defined. The imperfect work extraction protocol for qubits is depicted in \fig{fig:protocol}. 

As stated above, the geometric distance between  $\rho_\tth$ and the equilibrium state can be split into a coherence term and classical non-thermality term as per \eq{eq:avgrelE}.   These are shown  by the blue and red arrows in \fig{fig:protocol}, respectively. Below, we shall offer an intuitive quantification of coherence and classical non-thermality of the state $\rho_\tth$, named $\coh$ and $\noneq$ respectively, so that   $\coh(\rho_\tth)= \noneq(\rho_\tth) = 0$ if and only if $\rho_\tth = \tau_1$. These will be useful parameters in terms of which we may present our results later in the manuscript. 

The coherence of $\rho_\tth$ with respect to the Hamiltonian can be quantified by the minimum overlap between the eigenstates of $\rho_\tth$ and the eigenstates of $H^{(1)}$, i.e.
\be \label{eq:qubitcoh}
	\coh (\rho_\tth):=  \min_{k,l} |\<e_k|\tth_l\>|^2 = |\<e_+|\tth_-\>|^2 = \sin^2(\tth/2).
\ee
Hence $\coh(\rho_\tth) = 0$ for $\tth=0$, and it monotonically increases as $|\tth| \to \pi/2$, saturating at its maximum value of $\coh(\rho_{\pi/2})=1/2$.  The classical non-thermality of the qubit state $\rho_\tth$ compared to the thermal state $\tau_1$ for $H^{(1)}$ can be quantified by the logarithm of the ratio of ground state probabilities, i.e.
\be   \label{eq:qubitnoneq}
	\noneq (\rho_\tth) :=  \log {q_1 \over r_\tth}, 
\ee
where $q_1 :=\<e_-|\tau_1|e_-\>$ and $r_\tth = \<e_-|\rho_\tth|e_-\> = \<e_-|\eta_\tth|e_-\>$ are the  
ground state populations of  $\tau_1$ and $\rho_\tth$, respectively, see \fig{fig:protocol}. Hence $\noneq(\rho_\tth) = 0$ when $q_1 = r_\tth$, while a positive (negative) $\noneq (\rho_\tth)$ corresponds to a lower (higher) ground state population in $\rho_\tth$ than that of the thermal state $\tau_1$, corresponding to a down (up) red arrow in \fig{fig:protocol}.

\subsection*{Stochastic quantum trajectories} \label{sec:trajectories}

Working on the level of density matrices of the system during the protocol (see Fig.~\ref{fig:protocol} for the qubit example) limits the discussion of thermodynamic quantities to macroscopic expectation values only. In contrast, stochastic thermodynamics associates heat $Q (\Gamma)$, work $W (\Gamma)$ and entropy production $s_\irr (\Gamma)$ to individual microscopic trajectories $\Gamma$ forming the set of possible system evolutions \cite{Seifert2008,Sekimoto2010}. In this more detailed picture the macroscopic thermodynamic quantities $\avg{Q}, \avg{W}$ and $\avg{S}$ arise as weighted averages over these trajectories. In the quantum regime, quantum stochastic thermodynamics captures the set of possible trajectories that, in addition to classical trajectories, are determined by quantum coherences and non-thermal sources of stochasticity \cite{Manzano-fluctuation-quantum-maps,Alexia-measurement-thermodynamics, Manzano2017, Murashita-fluctuation-coherence-feedback, Grangier2018, Elouard-Book}. These trajectories consist of time-sequences of pure quantum states taken by an open system in a single run of an experiment.

 One way to experimentally `see'  quantum trajectories is by observing a sequence of stochastic outcomes of a generalized measurement performed on a system \cite{Exploring}. Immense experimental progress in the ability to measure quantum states with high efficiency has enabled the observation of individual jumps in photon number, and more recently the tracking of single quantum trajectories of superconducting qubits \cite{Gleyzes2007,Campagne-Ibarcq2016,Alonso2016,Murch2013}.  The natural set of quantum trajectories is a function of how the system is measured, and various quantum trajectory sets have been discussed in the literature each corresponding to different measurement setups: the so-called ``unravellings''  \cite{Carmichael2008,Gammelmark2013}.  Averaging the system's pure states over many experimental runs then gives back the density matrix describing the system's mixed state, whose evolution is governed by completely positive, trace preserving maps, also known as a quantum channel. Using the methods of quantum stochastic thermodynamics we here access a system's fluctuations in work, heat and entropy production, when quantum coherences are involved and irreversibility occurs. This allows us to expose the microscopic links between irreversibility and energetic exchanges in the quantum regime. 

We here use ``eigenstate trajectories'' that describe a system that travels through a sequence of eigenstates of its time-local density operators. Namely, the system is measured at instances in time $j = 1, 2, ... $ in the instantaneous eigenbases of the states $\rho_j$ that are assumed to be known, for example,  from a master equation that describes the open system dynamics. We note that this is an idealized scenario as in general one does not know what the density operators $\rho_j$ are and cannot guarantee to measure in the correct eigenbases. The eigenstate trajectories are analytically tractable, and provide a  convenient analytical tool to investigate the energetic footprints of irreversibility, as we will see below. 

\begin{figure*}[t!]
\includegraphics[trim=2cm 6cm 2cm 4cm,clip, width=0.7\textwidth]{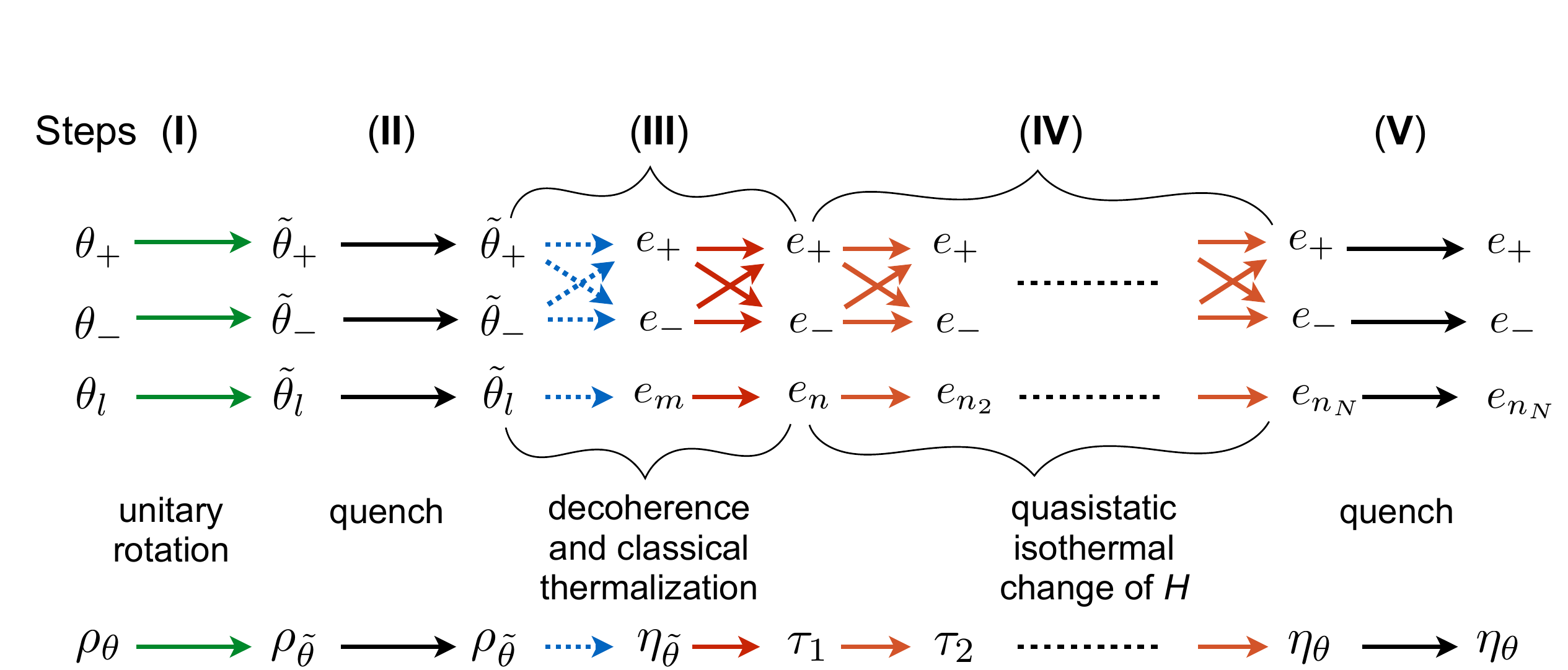}
\caption{\sf 
{\bf Pure-state qubit trajectories for the work extraction protocol.} 
Illustration of the evolution of the qubit during the work protocol on the trajectory level and on the density matrix level. The qubit's trajectories are deterministic during Steps {\bf (I)} (unitary, green arrows), {\bf (II)} (quench, black arrows), and {\bf (V)} (quench, black arrows), i.e. they take one state to a unique other state. In contrast, during the decoherence part in Step {\bf (III)} (blue dashed arrows) the qubit stochastically jumps from one of the states $\ket{\tth_\pm}$ to one of the energy eigenstates $\ket{e_\pm}$, thus losing any quantum coherence in an irreversible manner. During the classical thermalization part in Step {\bf (III)} (red arrows) the qubit stochastically jumps from one of the energy eigenstates to another energy eigenstate, thus losing any classical non-thermality in an irreversible manner. The qubit's trajectories during the classical quasistatic isothermal change of $H$ (Step {\bf (IV)}, orange arrows), are stochastic but reversible, due to infinitely small thermalizations taking place throughout. } \label{fig:trajectories}
\end{figure*}

The ensemble of trajectories $\{\Gamma \}$ taken by a quantum system  when undergoing the work extraction protocol outlined in the previous section can be broken up into trajectories for each of the Steps (see \fig{fig:trajectories} for the qubit example).  We will here focus on discussing the thermalization of the system  in Step {\bf (III)}, for which the initial density matrix $\tilde \rho$  can host coherences $D[\tilde \rho \| \tilde \eta] > 0$ and classical non-thermality $D[\tilde \eta \| \tau_1] > 0$ at the point when it is brought in contact with the thermal bath. The trajectories for the full protocol are detailed in the Methods.

The thermalization process in Step {\bf (III)} may be described by the quantum channel $\Lambda(\rho) := \tr\sub{\bb}[\vv (\rho \otimes \tau\sub{\bb}) \vv^\dagger]$ where $\tau\sub{\bb}:= \exp(- H_\bb / k_B T)/Z\sub{\bb}$ is the initial thermal state of the bath with Hamiltonian $H\sub{\bb}$ and partition function $Z\sub{\bb}$, and $\vv$ is a unitary operator that commutes with $H^{(1)} + H\sub{\bb}$. Hence $\Lambda$ is a thermal operation \cite{Perry2015,Lostaglio2016,Huei2018}. We further demand that $\Lambda$ is a fully thermalizing map, i.e.  $\Lambda(\rho) = \tau_1$ for all $\rho$. This map exists, for example, when the bath is chosen as an infinite ensemble of identical particles, each with the same Hamiltonian as the system, and with $\vv$ implementing a sequence of partial swaps between the system and each bath particle, or a full swap with just a single particle \cite{Ziman2010}.
Minimal trajectories for the thermalization process can now be constructed as $\GthreeSmin \equiv \ket{\tilde \psi_l} \mapsto \ket{e_n}$ ( see Fig.~\ref{fig:trajectories} for the specific case where the system is a qubit, with $|\tilde \psi_l\> \equiv |\tth_\pm\>$). The probability of this transfer to occur is 
$P\left( \GthreeSmin \right) 
= \<\tilde \psi_l| \tilde \rho |\tilde \psi_l\> \<e_n|\Lambda \left( \pr{\tilde \psi_l} \right) |e_n\>$,
which is obtained by first projectively measuring the system with respect to the eigenbasis $|\tilde \psi_l\>$ of $ \tilde \rho$, then applying the thermalization channel $\Lambda$, and finally measuring the system with respect to the eigenbasis $|e_n\>$ of $\tau_1$.
Since $\vv$ commutes with the total Hamiltonian while $\tau\sub{\bb}$ commutes with the bath Hamiltonian, it can be shown (see Theorem 1 in \cite{Mohammady2019a}) that $\<e_n|\Lambda \left( \pr{\tilde \psi_l} \right) |e_n\> = \sum_m |\<e_m|\tilde \psi_l\>|^2\<e_n|\Lambda \left(\pr{e_m}\right) |e_n\>$, where  $\ket{e_m}$ are eigenstates of the system Hamiltonian $H^{(1)}$. We may therefore ``augment'' our trajectories by projecting the system onto the energy basis $|e_m\>$ first  before letting it thermalize classically \cite{Elouard-Book}. 

The augmented trajectories are denoted $\GthreeS \equiv \ket{\tilde \psi_l} \mapsto  \ket{e_m} \mapsto  \ket{e_n}$, with probabilities
\be \label{eq:augmented-traj-foreward}
	P\left(\GthreeS \right) 
	= \<\tilde \psi_l|  \tilde \rho |\tilde \psi_l\> |\<e_m|\tilde \psi_l\>|^2 \<e_n|\tau_1|e_n\>.
\ee
It can be shown that the minimal trajectories $\GthreeSmin$ and the augmented trajectories $\GthreeS$ are thermodynamically equivalent, as they result in the same entropy production (see Methods for details).
However, the augmented trajectories have the benefit of naturally splitting into a  ``decoherence trajectory'' $\Gq \equiv \ket{\tilde \psi_l} \mapsto \ket{e_m}$, followed by a ``classical thermalization trajectory'' $\GclS \equiv \ket{e_m} \mapsto \ket{e_n}$, as depicted in  \fig{fig:trajectories} for the qubit case. 
Their probabilities to occur are
\be \label{eq:qtraj-probability}
	P \left(\Gq \right) 
	&=& \sum_{n} P\left(\GthreeS\right)
	= \<\tilde \psi_l| \tilde \rho |\tilde \psi_l\> |\<e_m|\tilde \psi_l\>|^2, \quad \quad
\ee
and 
\be \label{eq:cltraj-probability}
	P\left(\GclS \right) 
	&=& \sum_l P\left(\GthreeS \right)
 	= \<e_m| \tilde \eta|e_m\> \<e_n|\tau_1|e_n\>, \quad \quad
\ee
respectively which can be obtained as marginals of the probability distribution given by \eq{eq:augmented-traj-foreward} (see Methods for details). Here $ \Gq$ are the trajectories the system undertakes as it undergoes the decoherence process $ \tilde \rho \mapsto \tilde \eta$, while $\GclS$ are the trajectories that the system undertakes as it undergoes the classical thermalization process $\tilde \eta \mapsto \tau_1$. 

We note that while  \cite{Santos2019} also considered augmented trajectories to separate the quantum and classical contributions to the stochastic entropy production, these constituted of the initial and final energy eigenstates of the bath, together with initial and final eigenstates of the system, neither of which are  assumed to be energy eigenstates.  In our approach, the assumption that $\Lambda(\tilde \rho)$ is energy incoherent allows for the heat exchange of the process, in addition to the entropy production, to be split into a quantum and classical component, which we discuss below.

\subsection*{Stochastic quantum entropy production} \label{sec:ent}

Within quantum stochastic thermodynamics the entropy production along a quantum trajectory $\Gamma$ is
\be \label{eq:defn-entropy-production}
	s_\irr (\Gamma) :=  k_B \, \log {P(\Gamma) \over P^*(\Gamma^*)} ,
\ee
exposing the entropy production's microscopic origin as the imbalance between the probabilities $P(\Gamma)$ and $P^*(\Gamma^*)$ of a forward trajectory $\Gamma$ and its corresponding  backward trajectory $\Gamma^*$, respectively  \cite{Manzano2017, Elouard-Book}. The backward trajectory $\Gamma^*$ can be understood as the time-reversed sequence of eigenstates which constitute the forward trajectory $\Gamma$. In order to evaluate the probability for the backward trajectory, we consider the time-reversed process as one where the system and environment are initially in the compound state $\tau_1 \otimes \tau\sub{\bb}$, i.e.  the system starts in the average state that it took at the end of the forward process, while the bath is in thermal equilibrium.  On this initial product state, the time-reverse of the forward evolution of system and bath is applied, and projections are performed in reversed order into the forward eigenstates $\ket{\tilde \psi_l} $ and $\ket{e_m}$. This leads to Kraus operators given in \eqref{eq:backK} which describe the time-reversed trajectories, see Methods.  

We find that the stochastic entropy production for the thermalization  Step \textbf{(III)} can be expressed as
\be  \label{eq:Esplit}
	s_\irr \left(\Gthr\right) &=& \sqG + \sclG,
\ee
where we identify
\be \label{eq:quantum-entropy-production} 
	\sqG 
	&=& k_B \log {\langle \tilde \psi_l |  \tilde \rho |\tilde \psi_l \rangle   \over \langle e_m |  \tilde \eta | e_m \rangle} 
\ee
as the \emph{stochastic quantum entropy production}, and
\be \label{eq:classical-entropy-production}
	\sclG
	&=& k_B \log {\langle e_m |  \tilde \eta | e_m \rangle \over \langle e_m | \tau_1 | e_m \rangle}
\ee
as the \emph{stochastic classical entropy production}.
Since the probability of the augmented trajectories, $P\left(\GthreeS \right)$, gives $P\left(\Gq\right)$ and $P\left(\GclS\right)$ as marginals (see \eq{eq:qtraj-probability} and \eq{eq:cltraj-probability}), the average entropy production in Step \textbf{(III)} can also be split into an average quantum entropy production $\qent$, and an average classical entropy production, $\clent$. 
One finds, see  Methods, that each of these averages reduces to a relative entropy between two pairs of system states, 
\be \label{eq:aept1} 
	\qent 
	&=& \sum_{l, m} P\left(\Gq\right) \, \sqG 
	= k_B \, D[ \tilde \rho \|  \tilde \eta],  \quad \quad \\
	\label{eq:aept2} 
	\clent 
	&=& \sum_{m, n} P\left(\GclS\right) \, \sclG 
	= k_B \,  D[ \tilde \eta \| \tau_1] . \quad \quad
\ee
This shows that the relative entropies $D[ \tilde \rho \|  \tilde \eta]$ and $D[\tilde \eta \| \tau_1]$, which geometrically link density matrices, are physically meaningful as the average entropy productions associated with the evolution of the quantum system along ensembles of quantum trajectories. The two separate contributions to the entropy production arise because the system has two distinct non-equilibrium features, coherence with reference to the Hamiltonian, and classical non-thermality. Each is irreversibly removed when the system is brought into contact with the thermal bath and undergoes decoherence trajectories followed by classical thermalization trajectories. 

Finally, we show in Methods that the average entropy production for the full protocol reduces to $\qent + \clent$ in the limit where Step \textbf{(IV)} becomes a quasistatic process, i.e.  in this limit the average entropy production for the full protocol coincides with the average entropy production for the thermalization step alone.

\subsection*{Classical and quantum heat distributions} \label{sec:heat}

We now analyze the  energetic fluctuations of the quantum decoherence and classical thermalization trajectories, $\Gq$ and $\GclS$, respectively.  Since no external control is applied during these trajectories, such as a change of Hamiltonian, no work is done on the system and hence the energetic changes of the system consist entirely of heat. But since we identified two contributions to irreversibility, namely quantum decoherence and classical thermalization, it stands to reason that we should obtain two types of heat \cite{Alexia-measurement-thermodynamics, Alexia-fluctuation-engineered-reservoir}. 

The microscopic mechanisms associated with classical thermalization 
of the system with the bath are the quantum jumps from $\ket{e_m}$ to $\ket{e_n}$, which give rise to energetic fluctuations.  The heat the system \emph{absorbs} from the bath is
\be \label{eq:classical-heat}
		\Qcl\left( \GclS \right) = E_n^{(1)} -  E_m^{(1)} ,
\ee 
where $E_k^{(j)} := \<e_k| H^{(j)} |e_k\>$, which is the standard classical stochastic heat.  We note that  Step {\bf (IV)} also incurs classical heat, but we do not discuss this contribution here, as the stochastic thermodynamic description is well established for  heat exchanges during this classical quasistatic isothermal process \cite{Seifert2008, Sekimoto2010}.

\begin{figure}[t]
\includegraphics[ width=0.5\textwidth]{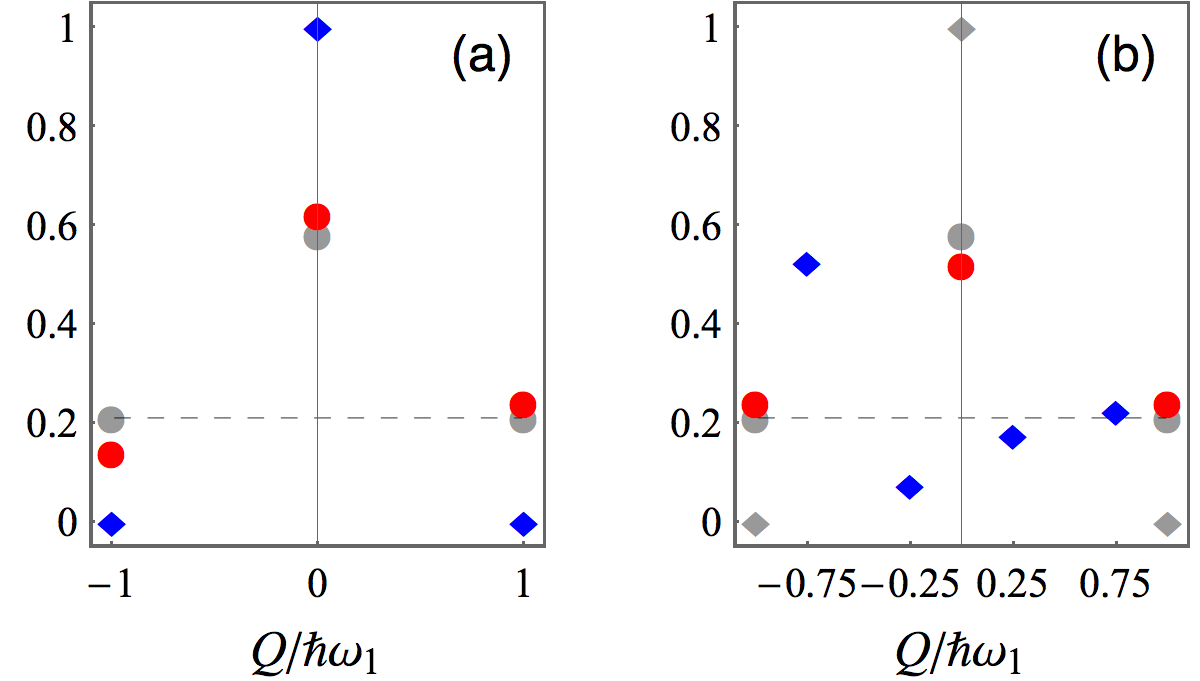} 
\caption{\sf {\bf Heat distributions for a qubit undergoing the thermalization Step {\bf (III)}}. 
Histograms of classical heat $\Qcl$ (red circles) and quantum heat $\Qq$ (blue squares) for
{\bf (a)} an initial state $\rho_\tth$ that hosts classical non-thermality: $\noneq (\rho_\tth) = \log (0.2/0.3)$ and $\coh (\rho_\tth) = 0$,  
and for {\bf (b)} an initial state $\rho_\tth$ that hosts quantum coherence: $\coh (\rho_\tth) = \sin^2(\pi/6)=1/4$ and $\noneq (\rho_\tth) = 0$. 
For comparison, grey circles and grey diamonds in both panels show the classical and quantum heat histograms, respectively, for when Step {\bf (III)} is fully reversible, i.e. $\rho_\tth = \tau_1$ and hence $\coh(\rho_\tth) = 0 = \noneq(\rho_\tth)$. Note that, even then the system can exchange heat with the bath leading to a classical heat distribution with non-zero but symmetrical values (dashed line) that give a zero average classical heat. 
In {\bf (a)} the only quantum heat value with non-zero probability is 0 (no quantum heat when thermalising a classical state), while in {\bf (b)} four non-trivial quantum heat values occur since  $\coh(\rho_\tth) \neq 0$.
\label{fig:Qdistribution}
} 
\end{figure}

On the other hand, the microscopic mechanisms associated with decoherence are the quantum jumps from $\ket{\tilde \psi_l}$ to $\ket{e_m}$, which give rise to energetic fluctuations of the system that are entirely quantum mechanical. 
The system's energy increase due to decoherence is    
\be \label{eq:quantum-heat}
	\QqG = E_m^{(1)}  - \bra{\tilde \psi_l} H^{(1)}\ket{\tilde \psi_l}.
\ee
It  has no classical counterpart and is hence referred to as \emph{quantum heat} \cite{Alexia-measurement-thermodynamics, Alexia-fluctuation-engineered-reservoir}.  Contrary to the classical stochastic heat which has fixed quantized values given by the Hamiltonian $H^{(1)}$ alone, the stochastic quantum heat's values vary as a function of the eigenbasis of the  state $\tilde \rho$.  When this state has no quantum coherences ($D[\tilde \rho \| \tilde \eta]=0$) the only realised value of the stochastic quantum heat is 0, i.e. in the absence of coherences, decoherence has no effect on the system's state and no energetic fluctuations result from it. Fluctuations of the quantum heat take place as soon as $D[\tilde \rho \| \tilde \eta] > 0$. Histograms of the classical stochastic heat $\Qcl$ and the quantum heat $\Qq$ for the qubit model are shown in Fig.~\ref{fig:Qdistribution}(a) and \ref{fig:Qdistribution}(b) for  states $\rho_\tth$ that have only classical non-thermality while $\coh = 0$, and states that have only coherences while $\noneq =0$, respectively.

Note that we were able to split the energetic changes of the thermalization process into \eq{eq:classical-heat} and \eq{eq:quantum-heat} by first augmenting the minimal trajectories  $\GthreeSmin$ to $\GthreeS$, and then splitting these into a decoherence trajectory $\Gq$ followed by a classical thermalization trajectory $\GclS$. While the minimal trajectories only consider transitions between the system's time-local eigenbases, and the projective measurements which realise them are therefore ``non-invasive'', the same is not true for the augmented trajectories which require a projective energy measurement on the system prior to thermalization, which destroys any coherence present.  Notwithstanding, since this energy measurement does not alter the stochastic entropy production   one can consider it as a ``virtual process'' that need not be actually performed. But to physically observe the quantum heat distribution would necessitate such an energy measurement, and then the source of the quantum heat originates from the projective energy measurement itself, and not from the thermal bath, as first discussed in \cite{Alexia-measurement-thermodynamics}.  

\subsection*{Heat footprints of classical and quantum irreversibility}  \label{sec:footprint}

We are now ready to discuss the energetic footprints of irreversibility in the quantum regime. The energetic footprints of classical entropy production during Step \textbf{(III)} are made immediately apparent from the stochastic equation (\ref{eq:classical-entropy-production}) which, in conjunction with the classical heat value given by \eq{eq:classical-heat}, can be re-expressed as 
\be
	\sclG &=& k_B \log {\langle e_m | \tilde \eta | e_m \rangle \over \langle e_n | \tau_1 | e_n \rangle} 
	- \frac{ \Qcl\left( \GclS \right)}{T}.  \quad \quad
\ee
When averaged over the classical thermalization trajectories $\GclS$, the above expression links the average absorbed  heat $\Qclavgwo$ to the average entropy production $\clentwo$ as
\be \label{eq:avg-cl-heat-footprint}
 	\clentwo = k_B (\SvN(\tau_1) - \SvN(\tilde \eta))  - \frac{\Qclavgwo}{T}  . \quad 
\ee
This thermodynamic equality, going back to Clausius, is the well-known energetic footprint of entropy production in the classical regime. It can be used to define the irreversibly dissipated heat,
\be 
	\avg{\qdiss}
	&:=& - \Qclavgwo +  T \DScl  
	= T  \clentwo = k_B T \, D[\tilde \eta \| \tau_1]   \geqslant 0 , \label{eq:qdiss}
\ee 
which is strictly positive when the entropy production $\clentwo$ is non-zero, which  arises when    forward and backwards probabilities of the process deviate, see \eqref{eq:defn-entropy-production}. In other words, the energetic footprint of non-zero $\avg{\qdiss}$ gives thermodynamic testament of the arrow of time.

\medskip

Meanwhile, the stochastic quantum entropy production $\sqG$  in \eq{eq:quantum-entropy-production} is given purely by a stochastic quantum entropy change and does not appear to involve any contributions from the stochastic quantum heat $\Qq$ whatsoever. When averaged over all quantum decoherence trajectories, the quantum heat in fact vanishes, see Methods,
\be  
	\Qqavg &=& 0,
\ee
while the average quantum entropy production can formally be rewritten as 
\be \label{eq:avg-q-heat-footprint}
 	\qentwo & = k_B (\SvN(\tilde \eta) - \SvN(\tilde \rho)) - \frac{1}{T}\cancelto{0}{\Qqavgwo}. \quad \quad 
\ee
This quantum thermodynamic equality shows that the energetic footprint of quantum entropy production, i.e. a fixed relationship between average heat absorption and average entropy production, is mute in the quantum regime.  This indicates a fundamental difference in how quantum and classical heat relate to the entropy production.

\bigskip

While \emph{prima faciae}   \eq{eq:avg-q-heat-footprint} seems to suggest that the quantum entropy production is completely dissociated from quantum heat, such a conclusion is premature. Indeed, on closer examination we discover that the average quantum entropy production $\qentwo$ is intimately linked with the variance in quantum heat, $\Var{\Qq}$, a quantity that has recently been connected to witnessing entanglement generation \cite{Elouard2018}. Specifically, we shall show that  $\qentwo=0$ is both necessary and sufficient for $\Var{\Qq}=0$, and for the special case of qubits, they are co-monotonic with energy-coherence of the system's state.   Before discussing this, let us first highlight that no such relationship exists between the average classical entropy production $\clent$ and the variance in classical heat, $\Var{\Qcl}$;   as shown in Methods, the variance in classical heat as the system thermalizes to $\tau_1$ takes the simple form of 
\begin{align}
\Var{\Qcl} = \Delta(H^{(1)}, \tilde \eta) + \Delta(H^{(1)}, \tau_1) \equiv \Delta(H^{(1)}, \tilde \rho) + \Delta(H^{(1)}, \tau_1),
\end{align}
where $\Delta(H, \rho):= \tr[H^2 \, \rho] - \tr[H \, \rho]^2$ is the variance of $H$ in state $\rho$. Clearly, $\clent=0$ is neither necessary nor sufficient for $\Var{\Qcl}=0$: (i) $\clent = 0$ if and only if $\tilde \eta = \tau_1$, whereas in such a case $\Var{\Qcl} = 2\Delta(H^{(1)}, \tau_1) \geqslant 0$ with equality being achieved only in the limit of zero temperature; (ii) $\Var{\Qcl}=0$ if and only if   $\Delta(H^{(1)}, \tilde \eta)= \Delta(H^{(1)}, \tau_1) = 0$. This means that both $\tilde \eta$ and $\tau_1$ only have support on  a single energy subspace of the Hamiltonian, such energy subspace of $\tau_1$ necessarily being the lowest one. However, if the subspace of $\tilde \eta$ is disjoint from that of $\tau_1$, then $\clent = k_B D[\tilde \eta \| \tau_1] = \infty$.

 As shown in Methods, the variance in quantum heat for the state $\tilde \rho = \sum_l p_l \Pi[\tilde \psi_l]$ decohering with respect to the Hamiltonian $H^{(1)}$ is the avarage variance of $H^{(1)}$ in the pure states $\ket{\tilde \psi_l}$, i.e. 
\be \label{eq:variance-qheat-equality}
	\Var{\Qq} = \sum_l p_l \, \Delta(H^{(1)}, \tilde \psi_l) \equiv \sum_l p_l \, I_\alpha(H^{(1)}, \tilde \psi_l),
\ee
where  $I_\alpha(H, \rho):= \tr[H^2  \, \rho] - \tr[H\, \rho^\alpha \, H  \,  \rho^{1-\alpha}]$ for $\alpha \in (0,1)$ is the set of Wigner-Yanase-Dyson skew informations of the observable $H$ in the state $\rho$ \cite{Wigner1963, Lieb1973, Yanagi2010}.  This variance in quantum heat obeys the inequalities
\be \label{eq:variance-qheat-inequality}
	\Delta(H^{(1)}, \tilde \rho) \geqslant \Var{\Qq} \geqslant I_\alpha(H^{(1)}, \tilde \rho),  
\ee
where the equalities in \eq{eq:variance-qheat-inequality} are saturated when $\tilde \rho$ is a pure state.  

\begin{figure}[t]
\includegraphics[width=0.86\textwidth]{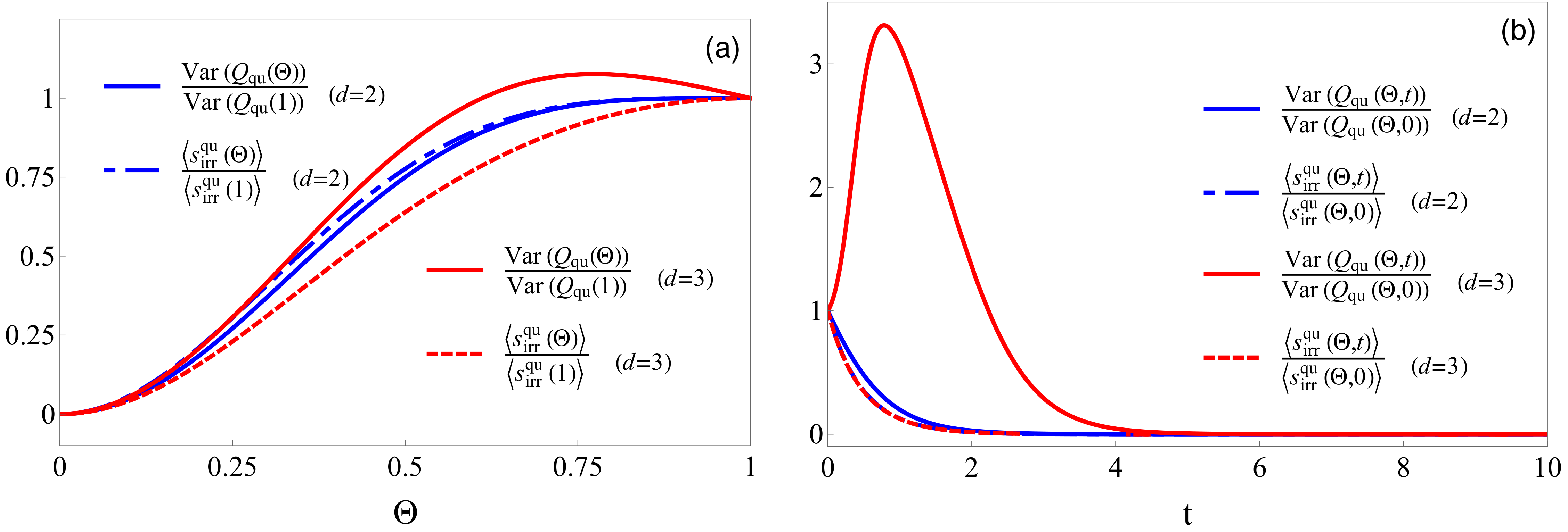}
\caption{\label{fig:cohc} {\bf Break-down of monotonic relationship between quantum entropy production and fluctuations in quantum heat for dimensions greater than two.} Here, we choose Hamiltonians with uniformly gapped spectra, i.e. $E_{k+1}^{(1)} - E_{k}^{(1)} = \hbar \omega_1$. The probability spectra for the states $\tilde \rho(\Theta)$ are chosen to be non-degenerate, but concentrated around $\ket{\psi_1^\Theta}$ and $\ket{\psi_d^\Theta}$. For $d=2$, $\bm{p} = (0.9, 0.1)$, while for $d=3$, $\bm{p} = (0.49, 0.04, 0.47)$.    
{\bf (a)}  Variance in quantum heat and average quantum entropy production as a function of $\Theta$ defined in  \eq{eq:Utheta}. For $d=2$, both $\Var{Q_\mathrm{qu}(\Theta)}$ and $\qenttheta$ monotonically increase as $\Theta \to 1$. For $d=3$, however, while $\qenttheta$ monotonically increases with $\Theta$,  $\Var{Q_\mathrm{qu}(\Theta)}$ takes a maximum value at $\Theta \approx 0.8$, after which it decreases.   {\bf (b)} Here we choose the initial states $\tilde \rho(\Theta)$ with $\Theta = 0.3$, and evaluate $\Var{Q_\mathrm{qu}(\Theta,t)}$ and $\qentthetat$ for the states $e^{t \lo} (\tilde \rho(\Theta))$ with $\lo$ defined in \eq{eq:dephasing-lindblad}. For $d=2$, both $\Var{Q_\mathrm{qu}(\Theta,t)}$ and $\qentthetat$ monotonically decrease with $t$, while for $d=3$, $\Var{Q_\mathrm{qu}(\Theta,t)}$ takes its maximum value at $t \approx 1$.  } 
\end{figure}

Both $I_\alpha(H^{(1)}, \tilde \rho)$ and $\qentwo/k_B= D[\tilde \rho \| \tilde \eta]$ quantify the asymmetry of the state $\tilde \rho$ with reference to the Hamiltonian $H^{(1)}$, and are thus linked with the resource theory of asymmetry \cite{Vaccaro2008a, Ahmadi2013b, Baumgratz2014,  Marvian2014, Girolami2014,  Takagi2018}. Specifically, both $I_\alpha(H^{(1)}, \tilde \rho)$ and $D[\tilde \rho \| \tilde \eta]$ vanish if and only if $\tilde \rho$ commutes with $H^{(1)}$, and monotonically decrease under Hamiltonian-covariant quantum channels, i.e. quantum channels $\e$ which satisfy $\e(e^{-i t H^{(1)}} \rho e^{i t H^{(1)}} ) = e^{-i t H^{(1)}} \e (\rho) e^{i t H^{(1)}}$ for all $t \in \mathds{Re}$ and $\rho$.   Therefore, by \eq{eq:variance-qheat-equality} we  conclude that the average quantum entropy production vanishes if and only if the variance in quantum heat vanishes. Additionally,  given a pair of quantum states $\tilde \rho_1$ and $\tilde \rho_2 = \e (\tilde \rho_1)$, then: (a) the average quantum entropy production as $\tilde \rho_1$ decoheres to $\tilde \eta_1$ is no smaller than that obtained when  $\tilde \rho_2$ decoheres to $\tilde \eta_2$; and (b) by \eq{eq:variance-qheat-inequality}, the \emph{lower bound} to the quantum heat variance as $\tilde \rho_1$ decoheres to $\tilde \eta_1$ is no smaller than that obtained when  $\tilde \rho_2$ decoheres to $\tilde \eta_2$. Of course, this observation still allows for the existence of a pair of states  $\tilde \rho_1$ and $\tilde \rho_2 = \e(\tilde \rho_1)$ such that the average quantum entropy production of the former exceeds that of the latter, while the fluctuations in quantum heat of the latter exceeds that of the former.  In what follows we shall show that, surprisingly,   in the special case of qubits, i.e. $d=2$, the  fluctuations in quantum heat are  monotonic with the average quantum entropy production. This link is two-fold: (i) for two states with the same probability spectrum, but different eigenbases, the average quantum entropy production and the variance in quantum heat are monotonically increasing with the ``energy coherence'' of the eiganbasis; (ii) both the average quantum entropy production and the variance in quantum heat monotonically decrease under the action of Hamiltonian-covariant channels that are a combination of dephasing and depolarization. Both of these necessary links break down for higher dimensions,  which we illustrate with a simple counter example for $d=3$, see \fig{fig:cohc}.

Let us first consider how $\Var{\Qq}$ and $\qentwo$  are affected by the relationship between the eigenbasis of the quantum state $\tilde \rho$, and the eigenbasis of the Hamiltonian $H^{(1)}$.  Specifically, we shall consider a family of quantum states  $\tilde \rho(\Theta) := \uu(\Theta)  \tilde \rho \,  \uu^\dagger(\Theta)$ for $\Theta \in [0,1]$, where $\tilde \rho = \sum_l p_l \Pi[e_l]$ commutes with the Hamiltonian,  with the one-parameter unitary operator
\begin{align}\label{eq:Utheta}
\uu : [0,1] \ni \Theta \mapsto \exp (\Theta \log{\mathcal{F}})
\end{align}
being generated by  the discrete quantum Fourier transform  \cite{Vourdas2004} $\mathcal{F}$ defined as
\begin{align}\label{eq:MUB}
\mathcal{F} : \ket{e_l} \mapsto  \ket{\xi_l}  := \frac{1}{\sqrt{d}} \sum_{k=1}^d e^{2\pi i (l-1)(k-1)} \ket{e_k}. 
\end{align}
It is simple to verify that  $\{\ket{e_k}\}$ and $\{\ket{\xi_l} \}$ are a pair of mutually unbiased bases, with the energy coherence of $\{\ket{\xi_l} \}$ taking the maximum value of $\coh:= \min_{k,l} |\<e_k|\xi_l\>|^2 = 1/d$.  
We shall denote the eigenbasis of $\tilde \rho(\Theta)$ as $\mathscr{B}(\Theta):= \{\ket{ \psi_l^\Theta} \equiv \uu(\Theta) \ket{e_l}\}$, and the probability spectrum of $\tilde \rho(\Theta)$ and $\tilde \eta(\Theta) := \sum_k \Pi[e_k] \tilde \rho(\Theta) \Pi[e_k]$ as $\bm{p} := (p_l)_l$ and $\bm{r}(\Theta)$, respectively.    The Hamiltonian $H^{(1)}$ will map $\mathscr{B}(\Theta)$ to the symmetric doubly stochastic matrix $\bm{M}(\Theta)$, which has the matrix elements $M_{k,l}^{(\Theta)} :=  |\<e_k|\uu(\Theta)|e_l\>|^2$. Both the quantum heat variance and average quantum entropy production can be computed by knowledge of these matrix elements: the quantum entropy production can be computed as $\qentwo = k_B \left(\SvN(\tilde \eta(\Theta)) - \SvN(\tilde \rho(\Theta)) \right) \equiv k_B \left(\mathscr{H}(\bm{r}(\Theta)) - \mathscr{H}(\bm{p}) \right)$, where $\mathscr{H}$ denotes the Shannon entropy, and $\bm{r}(\Theta) = \bm{M}(\Theta) \bm{p}$;  the variance in quantum heat can be computed, as \eq{eq:variance-qheat-equality}, by 
\be\label{eq:variance-Theta}
	\Delta(H^{(1)}, \psi_l^\Theta) = \sum_{k=1}^d M_{k,l}^{(\Theta)} \bigg(E_k^{(1)} 
	- \sum_{k'=1}^d M_{k',l}^{(\Theta)} E^{(1)}_{k'} \bigg)^2.
\ee

When $d=2$,  we have $M_{k\ne l,l}^{(\Theta)} =  \frac{1}{2}\sin^2(\Theta \pi/2) \equiv \coh$ and $M_{l,l}^{(\Theta)} = 1 - \frac{1}{2}\sin^2(\Theta \pi/2) \equiv 1-\coh$, where we recall that  $\coh := \sin^2(\tth/2)$ for $\tth \in [-\pi/2, \pi/2]$ (see \eq{eq:qubitcoh}). Consequently, by \eq{eq:variance-qheat-equality} and \eq{eq:variance-Theta}, the variance in quantum heat takes the simple form of 
\be \label{eq:avg-q-heat-footprint2}
	\Var{\Qq} 
	=  \left(\hbar \omega_1 \right)^2 \left(\coh - \coh^2\right) \equiv \frac{\left(\hbar \omega_1 \right)^2}{4} \sin^2(\tth) ,
\ee
for all probability spectrums $\bm{p}$  (see  Methods for details). As such, $\Var{\Qq}$ vanishes when $\Theta = 0 = \coh$, and monotonically increases with $\Theta$, or equivalently with $\coh$, for all $\bm{p}$ and Hamiltonians $H^{(1)}$. As for the entropy production, we note that $D[\tilde \rho(0) \| \tilde \eta(0)]=0$, and that for any $\Theta_2 \geqslant \Theta_1$, there exists a $\Theta'$ such that $\bm{M}(\Theta_2) = \bm{M}(\Theta') \bm{M}(\Theta_1)$. Due to the properties of doubly stochastic matrices and majorization, this is a sufficient condition for $\mathscr{H} (\bm{r}(\Theta_2) ) - \mathscr{H} (\bm{r}(\Theta_1) ) \geqslant 0$, which implies that $\qentwo$ also monotonically increases with $\Theta$, or equivalently with $\coh$, for all $\bm{p}$ and $H^{(1)}$\cite{Sherman1952,Bhatia1997,Li2013a,Ljubenovic2015}. The co-monotonic relationship between $\Var{\Qq}$ and $\qentwo$ with $\Theta$ for qubits is demonstrated in \fig{fig:cohc}{(a)}. Conversely, when $d=3$ we see that while $\qentwo$ monotonically increases with $\Theta$, the same is not necessarily true for $\Var{\Qq}$ which in this instance takes a maximum value at $\Theta \approx 0.8$. Here, we have chosen the Hamiltonian to have a uniform spectral gap, i.e. $E_{k+1}^{(1)} - E_{k}^{(1)} = \hbar \omega_1$, with the non-degenerate probability spectrum $\bm{p}$ concentrated around $\ket{\psi_1^\Theta}$ and $\ket{\psi_3^\Theta}$. The reason for this is that $\Delta(H^{(1)}, \psi_l^\Theta)$ is maximised when the probability distribution $(M_{k,l}^{(\Theta)})_k$ is concentrated around the smallest and largest energy eigenvalues $E_1^{(1)}$ and $E_d^{(1)}$ \cite{Bhatia-Variance}. While this is certainly achieved at $\Theta = 1$ for qubits, this is no longer the case for larger systems, where $(M_{k,l}^{(1)})_k= (1/d, \dots, 1/d)$.

Next, we consider how  $\Var{\Qq}$ and $\qentwo$ are affected by a Hamiltonian-covariant quantum channel $\e$. As stated previously, $\qentwo$ is known to monotonically decrease with  applications of $\e$, i.e. for $ \rho_2 = \e( \rho_1)$,  $D[ \rho_1 \|  \eta_1] \geqslant D[ \rho_2 \|  \eta_2]$. Moreover, as shown in  Methods, so long as $\e$ is a convex combination of pure dephasing with respect to the Hamiltonian eigenbasis, and a depolarization channel which takes the system to the complete mixture, then for qubits $\coh(\rho) \geqslant \coh(\e(\rho))$ for all $\rho$. Consequently, by 
\eq{eq:avg-q-heat-footprint2} the fluctuations in quantum heat for $\e(\rho)$ will be smaller than that of $\rho$. We demonstrate this in \fig{fig:cohc}{(b)} for the Hamiltonian-covariant, Markovian dephasing channels $\e(\rho) = e^{t \lo}(\rho)$, where 
\begin{align}\label{eq:dephasing-lindblad}
\lo(\rho) = \sum_k \pr{e_k}\rho \pr{e_k} - \frac{1}{2} \sum_k\bigg( \pr{e_k} \rho + \rho \pr{e_k} \bigg).
\end{align}
It is simple to verify that $\<e_k | e^{t \lo} (\rho) |e_l\> = e^{- t (1 - \delta_{l,k})} \<e_k| \rho |e_l\>$, and so $e^{-i H t'}e^{t \lo} (\rho) e^{i H t'} = e^{t \lo} (e^{-i H t'}\rho e^{i H t'})$. As can be seen, for $d=2$ both $\Var{\Qq}$ and $\qentwo$ monotonically decrease with $t$. For $d=3$, however, while  $\qentwo$ monotonically decreases with $t$, $\Var{\Qq}$  does not.

\begin{figure}[t]
\includegraphics[ width=0.8\textwidth]{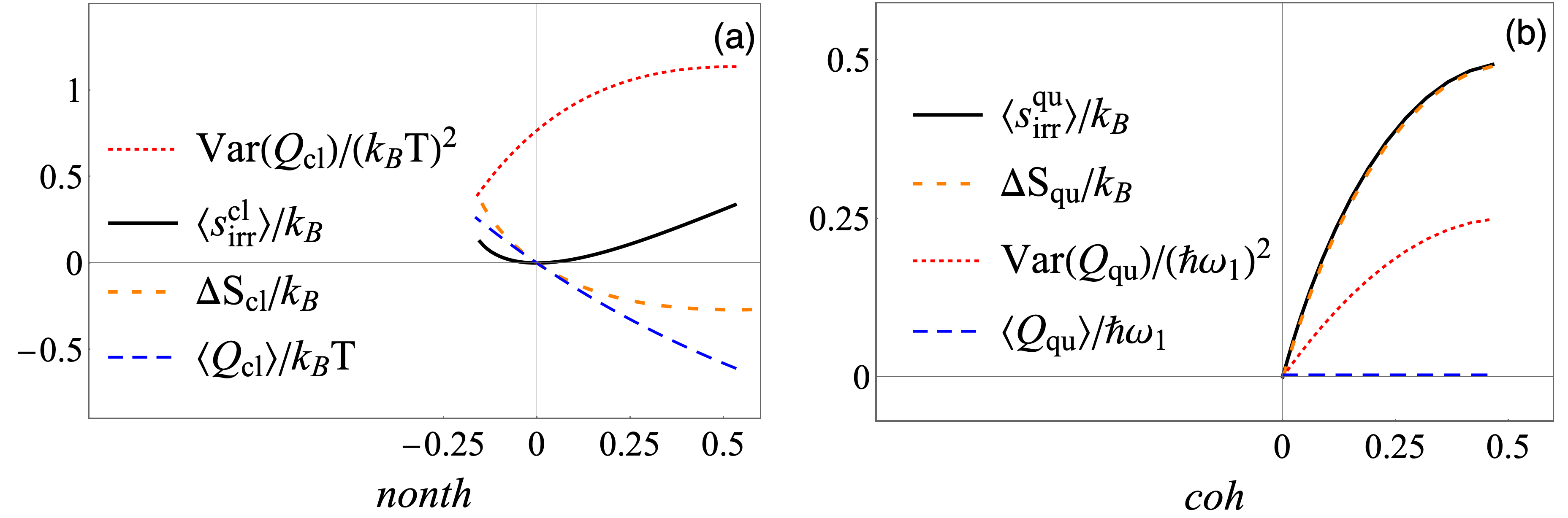} 
\caption{\sf 
{\bf Heat footprints of irreversibility for a qubit during Step {\textbf (III)}.}
{\bf (a)} Positive (negative) non-thermality $\noneq$ corresponds to a lower (higher) ground state population in $\eta_\tth$ than that of the thermal state $\tau_1$.
Qubit spacing vs thermal energy ($\hbar \omega_1/k_B T$) is here set such that $q_1 =\<e_-|\tau_1|e_-\> = 0.85$ while $p \in [0.5,1]$.  Classical entropy production $\clentwo$ plus the absorbed heat divided by the temperature, $\Qclavgwo/T$, gives the entropy change $\DScl$ for any classical non-thermality parameter $\noneq (\eta_\tth)$. At $\noneq=0$,  $\clentwo =0$ while the variance in classical heat,  $\Var{\Qcl}$, is strictly positive. Moreover, as   $\noneq$ grows more negative,  $\clentwo$ increases while $\Var{\Qcl}$ decreases. This demonstrates that the two quantities have no connection. 
\quad 
{\bf (b)} $\coh=0$ implies that  $\ket{\tth_\pm}$ are energy eigenstates, while as $\coh \to 0.5$, $\ket{\tth_\pm}$ become equal superpositions of energy eigenstates.
Initial state mixing probability is here set to $p=0.95$ while $\tth \in [0,\pi/2]$. Quantum entropy production $\qentwo$ plus zero average quantum heat $\Qqavgwo$ equals the entropy change $\DSqu$ for any quantum coherence parameter $\coh$ of initial states $\rho_\tth$. Also shown is the quantum heat variance $\Var{\Qq}$ in natural units $(\hbar \omega_1)^2$. Both, $\Var{\Qq}$ and $\qentwo$, increase monotonously as $\coh$ tends to its maximum value of $0.5$. } \label{fig:CQfluc}
\end{figure}

\bigskip

For the qubit case, Fig.~\ref{fig:CQfluc} puts in perspective the two drastically different energetic footprints of irreversibility in the classical and quantum regime.  On the well-known classical side, see Fig.~\ref{fig:CQfluc}a, the average entropy production $\clentwo$ is equal to the difference between the fixed entropy change $\DScl$ associated with the transfer $\eta_\tth \to \tau_1$, and an absorbed heat $\Qclavgwo$ when this transfer is achieved by an irreversible thermalization process, divided by the temperature $T$. The classical heat footprint $\Qclavgwo$ scales as the thermal energy $k_B T$, an energy scale set by the temperature of the bath that thermalizes the qubit. The more non-thermal the initial (diagonal) qubit state $\eta_\tth$ is, the more irreversibility will occur during its thermalization.  Hence the classical entropy production $\clentwo$ increases as the classical non-thermality parameter $\noneq (\eta_\tth)$ deviates from 0. Moreover, $\Var{\Qcl}$ is dissociated from $\clentwo$, since as $\noneq (\eta_\tth)$ approaches zero from below, $\clentwo$  becomes vanishingly small, while  $\Var{\Qcl}$ grows larger.

On the quantum side, see Fig.~\ref{fig:CQfluc}b, the average entropy production $\qentwo$ equals the entropy change $\DSqu = k_B (\SvN(\eta_\tth) - \SvN(\rho_\tth))$ associated with the decoherence $\rho_\tth \to \eta_\tth$ and does not link to an absorbed quantum heat $\Qqavgwo$, as this is always zero. However, both $\qentwo$ and the quantum heat fluctuations $\Var{\Qq}$ vanish when $\coh (\rho_\tth)=0$, and monotonously increase with  $\coh (\rho_\tth)$, showing the implicit link between quantum entropy production and quantum heat for qubits. This behaviour differs markedly from the classical counterpart.   Finally, we remark that unlike the classical case, the heat footprint does not scale with temperature but with the system energy gap, here $\hbar \omega_1$, an energy scale set by the quantum character of the system rather than the thermodynamics implied by the bath.

\subsection*{Fundamental bounds for work extraction}  \label{sec:workext}

\begin{figure}[!htb]
\includegraphics[trim=0cm 0cm 0cm 0cm,clip, width=0.49\textwidth]{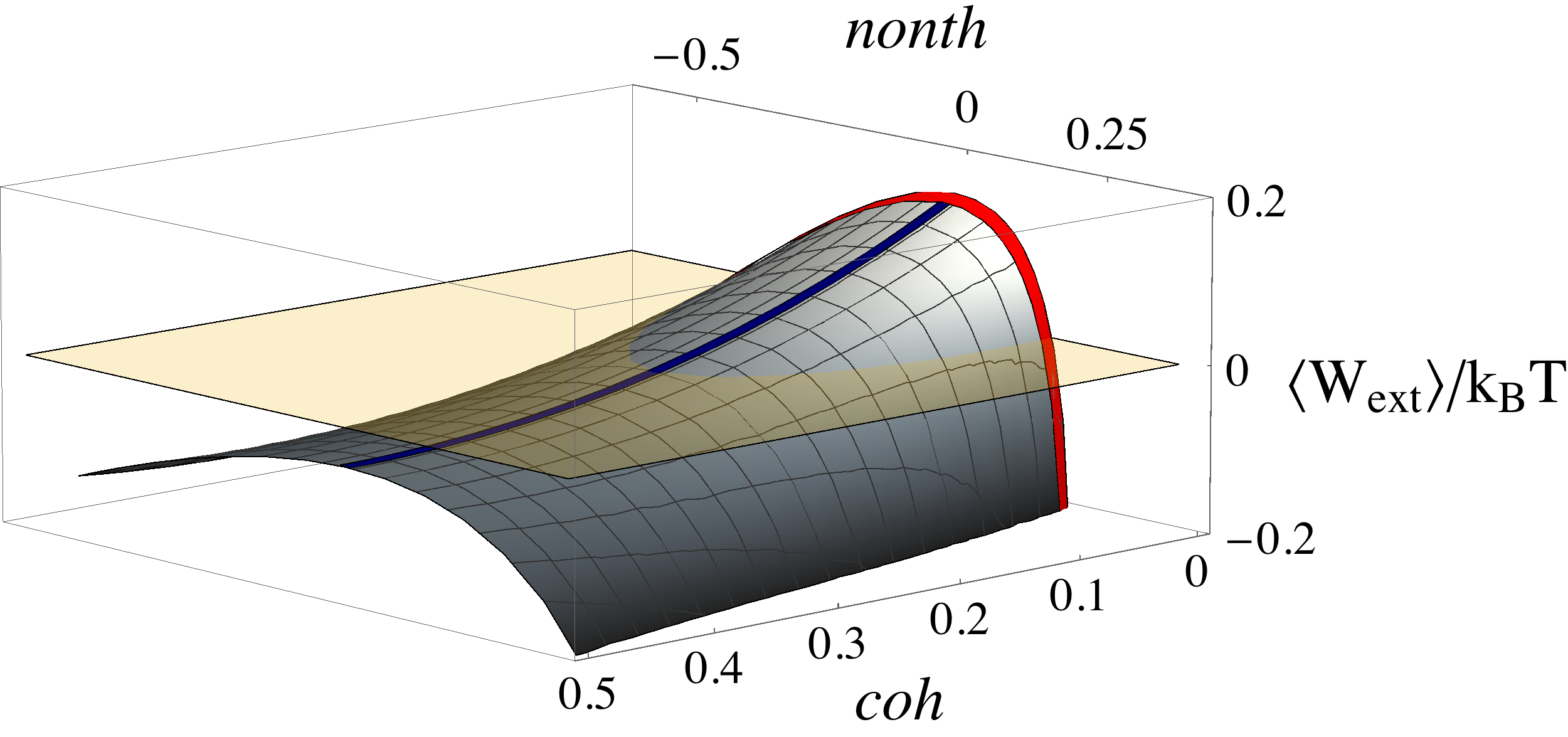}
\caption{
\sf {\bf Average work extraction from a qubit as a function of $\coh$ and $\noneq$.} 
Work (grey) for the full protocol is optimal when neither quantum coherence nor classical non-thermality is present, i.e. $\coh =0 = \noneq$, and the protocol is run reversibly \cite{Anders-coherence-thermo}. $\avg{\wext}$ decreases monotonously with increasing $\coh (\rho_\tth)$ (blue line for $\noneq=0$) and increasing and decreasing $\noneq(\rho_\tth)$  (red line for $\coh =0$). At large deviations from the reversible protocol, $\avg{\wext}$ becomes negative (crosses yellow plane at zero) and work would need to be invested to run the protocol. Parameter choices for initial qubit state $\rho_\theta$ are $p=0.8$ and $\theta= \pi/3$. 
}
\label{fig:work}
\end{figure}

Finally, we check the validity of the work footprint of entropy production, Eq.~\eqref{wext}, in the quantum regime. From the stochastic first law of thermodynamics, we observe that for each trajectory $\Gamma$ of the full protocol (see Methods for details) the stochastic extracted work is
\be \label{eq:stochastic work}
\wext(\Gamma) = \Delta U^{\rm prot}(\Gamma) + \Qq(\Gamma^\mathrm{q}) + \Qcl(\Gamma^\mathrm{cl}) + \Qcl^{\bm{\mathrm{(IV)}}}(\Gamma^{\bm{\mathrm{(IV)}}}),
\ee
where  $\Delta U^{\rm prot}(\Gamma) := \tr[H^{(0)}(\Pi[\psi_l] - \Pi[e_{n_N}])]$ is the decrease in internal energy along the trajectory $\Gamma$ for the full protocol; $\Qq(\Gamma^\mathrm{q})$ and $\Qcl(\Gamma^\mathrm{cl})$ are the quantum and classical heat absorbed during the thermalization process in Step {\bf{(III)}}; and $\Qcl^{\bm{\mathrm{(IV)}}}(\Gamma^{\bm{\mathrm{(IV)}}})
$ is the heat absorbed during the quasistatic process of Step {\bf{(IV)}}. Since $\avg{\Delta U^{\rm prot}} = \tr[H^{(0)}(\rho - \eta )] = 0$, while $\avg{ \Qq} = \tr[H^{(1)}(\tilde \eta - \tilde \rho )] = 0$, the average extracted work reduces to 
\be \label{eq:workaverage}
 	\avg{\wext}
	&=&   \Qclavg+ \Qivavg 
	=  -T \clentwo +  T \DScl  + T \DSFour .  
\ee
Here we have assumed quasistatic isothermal trajectories $\Gfour$ in Step {\bf (IV)} with $\siv =0$ and  thus
\be
	\Qivavg = T \, \DSFour =  T \, k_B (\SvN(\eta) - \SvN(\tau_1)). \nonumber
\ee
Substituting the entropy change across the entire protocol
\be
	\Delta S^{\rm prot}  = \DSqu + \DScl  + \DSFour,  \nonumber
\ee
and using $\Delta F^{\rm prot} = - T \Delta S^{\rm prot} = - k_B T (\SvN(\eta)- \SvN(\rho))$ since $\avg{\Delta U^{\rm prot}} =0$, 
the result is
\be  \label{eq:workfootprint}
	\avg{\wext}	&=& -  \Delta F^{\rm prot} - T \, (\clentwo + \qentwo).
\ee

Clearly, the optimum work value $- \Delta F^{\rm prot}$ is obtained when neither classical nor quantum entropy production are present and the process is run fully reversibly, as discussed in Ref.~\cite{Anders-coherence-thermo}. Equation \eqref{eq:workfootprint} now shows how the work is reduced when irreversible steps are included. It is evident that the classical and quantum entropy productions, $\clentwo$ and $\qentwo$, limit work extraction in a completely symmetrical manner and when these two contributions are combined \eq{eq:workfootprint} becomes identical to the well-known work-footprint of irreversibility, captured by \eq{wext}. This  footprint is shown in \fig{fig:work} for the qubit model, where $\avg{\wext}$ is plotted as a function of the two parameters that give rise to irreversibility, the quantum coherence $\coh$ and classical non-thermality $\noneq$ of the state $\rho_\tth$ before thermal contact.

While work extraction is mathematically limited in a symmetrical manner, the physical mechanism is drastically different depending on if the irreversibility of the protocol is of classical or of quantum nature. In the classical regime the irreversibly dissipated heat $\avg{\qdiss}$ is the physical cause of non-optimal work extraction and exactly compensates the non-recoverable work, i.e. the term $ T \, \clentwo = \avg{\qdiss}$ in \eq{eq:workfootprint}. This energetic footprint of irreversibility equals the average energy change of the qubit during the irreversible thermalization step. But the quantum decoherence step does not give rise to any average energy change - the work extraction is here reduced solely because the system \emph{entropy} increases, reducing the extracted work  by a proportional amount $ T \, \qentwo = T \, \DSqu$. 

\medskip

To conclude, when a quantum system loses its energetic coherences in a perfectly reversible manner, such as during a quasistatic thermodynamic protocol with a bath at temperature $T$, the energetic footprint is coherence work \cite{Anders-coherence-thermo} while no quantum heat occurs. On the other hand, when a quantum system loses its energetic coherences in a fully irreversible manner, such as during a quantum measurement, the energetic footprint is quantum heat \cite{Alexia-measurement-thermodynamics} while no coherence work occurs.  We here found that when a quantum system loses its energetic coherences in a partially reversible process, see Fig.~\ref{fig:protocol}, then the coherence work is in general non-zero, see Fig.~\ref{fig:CQfluc}, albeit reduced from the reversible case by a term proportional to the irreversible (quantum) entropy production, while the quantum heat distribution is also non-zero, see Fig.~\ref{fig:Qdistribution}(b). Surprisingly, it turned out that these two energetic footprints of irreversibility are not linked through entropy production in the same way as in classical physics.

\section*{Discussion}

The notion of irreversibility, and how it affects heat and work exchanges, is the core theme of thermodynamics. This paper brings  together several strands of recent research in quantum thermodynamics, including stochastic thermodynamics and quantum work extraction protocols, to provide a comprehensive picture of when irreversibility arises in the quantum regime and details the ensuing energetic footprints of irreversibility. 
Specifically, we have shown that the geometric entropy production as a quantum system in state $\tilde \rho$ thermalizes to $\tau_1$,  $k_B \, D[\tilde \rho \| \tau_1]$, which can be calculated using density matrices, can be understood as arising from the time-reversal asymmetry of quantum stochastic trajectories, \eq{eq:aept1} and \eq{eq:aept2}, in a similar way to classical stochastic thermodynamics.  In addition, the quantum eigenstate trajectories allowed for a detailed assessment of work and heat exchanges of a quantum system that can host coherences. While reversible work extraction from quantum coherences has been found \cite{Anders-coherence-thermo}  to give an ``average'' work of $\avg{\wext}^{\rm rev} = -  \Delta F^{\rm prot} $, no distribution of work was provided with respect to which $\avg{\wext}^{\rm rev}$ is an ``average''. Here we showed that quantum trajectories naturally give rise to heat as well as work distributions,  for which moments, such as the work ``average'', can be readily calculated.  By here including irreversible steps in the work extraction protocol, the reduction of work due to irreversibility has been quantified in \eq{eq:workfootprint}. Understanding how imperfect experimental control -- which leaves either quantum coherences, or classical non-thermality, or both present in a quantum system before thermal contact -- reduces work extraction is important for identifying experimental protocols that are optimal within realistic technical constraints. 

While the first moments of heat and work coincide with the values obtained on the density matrix level, the trajectories approach allows access to higher moments. This proved insightful for the discussion of the footprint of quantum irreversibility. We found that  the average classical  entropy production is linked to the surplus of dissipated  heat, see \eq{eq:qdiss},  which is fully analogous to the classical regime, see \eq{Qirr}. Conversely, no such link can be made in regards to quantum entropy production, see \eq{eq:avg-q-heat-footprint}. Instead, we show that the quantum entropy production is linked with the fluctuations in quantum heat. Specifically, we show that the average quantum entropy production vanishes if and only if the variance in quantum heat vanishes, while both the average quantum entropy production and the lower bounds to the variance in quantum heat monotonically decrease under Hamiltonian-covariant channels. In the specific case of qubits, we further show that: (i) for a family of states with the same spectrum but different eigenbases, both the  fluctuations in quantum heat and the average quantum entropy production monotonically increase with the energy coherence of the eigenbasis; (ii) both the fluctuations in quantum heat and the average quantum entropy production monotonically decrease under the action of Hamiltonian-covariant channels that are a mixture of pure dephasing and depolarization.   For higher dimensions, however, this necessary link breaks down in general.  We note that a comparable link does not exist in the classical regime where a vanishing classical entropy production is neither necessary nor sufficient for a vanishing variance in classical heat, and even for qubits the two quantities have no monotonic relationship.

It would be interesting to see if the same conclusions hold true when the eigenstate trajectories are replaced by experimentally measured trajectories and their probabilities, for which the analysis presented here can be implemented in an analogous manner. Another open problem is to establish a unique measure of the fluctuations in quantum heat for  degenerate states. It is known that if a quantum state has degenerate eigenvalues, then it offers infinitely many eigenstate decompositions, and hence the variance in quantum heat as quantified by \eq{eq:variance-qheat-equality} will not be uniquely defined by the quantum state alone.  While the lower and upper bounds in \eq{eq:variance-qheat-inequality} are independent of such an eigenstate decomposition, it would be interesting to introduce an operational procedure for measuring the fluctuations in quantum heat which are independent of the eigenstate decomposition of the system's state.

\medskip

\section*{Methods} 

In this section we provide detailed technical calculations for our main results, presented in the main text above. First, we describe the eigenstate trajectories for the full work extraction protocol, and the resulting entropy productions; Next we evaluate the variances in quantum and classical  heat as a  quantum system thermalizes, both for general $d$-dimensional systems and for qubits;  Finally we show that the energy coherence for all qubit states decreases under quantum channels that are a convex combination of dephasing with respect to the energy eigenbasis, and depolarization to the complete mixture.

\subsection*{Trajectories for the full work extraction protocol} \label{app:step-traj}

We now introduce the full trajectories of the protocol, with expressions for their probabilities, and evaluate the stochastic entropy production associated with each trajectory. We shall show that  the full  entropy production can be split into entropy production terms associated for each step. Next, we show that the average entropy production for the full protocol reduces to the average entropy production for Step \textbf{(III)} in the limit that the evolution in Step \textbf{(IV)} becomes quasistatic.

\,

Recall that the work extraction protcol can be split as follows. Step \textbf{(I)}: unitary evolution $\rho \mapsto \tilde \rho$;  Step \textbf{(II)}: Hamiltonian quench $H^{(0)} \mapsto H^{(1)}$; Step \textbf{(III)}: decoherence $\tilde \rho \mapsto \tilde \eta$ followed by classical thermalization $\tilde \eta \mapsto \tau_1$; Step \textbf{(IV)}: quasistatic evolution $\tau_1 \mapsto \dots \mapsto \tau_N \equiv \eta$; and Step \textbf{(V)}: Hamiltonian quench $H^{(N)} \mapsto H^{(0)}$. Since Steps \textbf{(II)} and \textbf{(V)} are only Hamiltonian quenches, and do not alter the state, we shall not include these when constructing our trajectories. 

Each thermalization process that the system undertakes is described by the channels $\Lambda_i : \rho \mapsto \tr\sub{\bb_i}[\vv_i (\rho \otimes \tau\sub{\bb_i}) \vv_i^\dagger]$, where $\bb_1 \equiv \bb$ and $\vv_1 \equiv \vv$ are the bath and unitary used in Step \textbf{(III)}, while $\bb_2,\dots, \bb_N$ and $\vv_2, \dots, \vv_N$ are the baths and unitaries used in Step \textbf{(IV)}. We shall decompose each thermalization channel into their Kraus operators $K_{\mu_i, \nu_i}:= \sqrt{\<\mu_i|\tau\sub{\bb_i}|\mu_i\>} \<\nu_i|\vv_i|\mu_i\>$, where $\ket{\mu_i}$ and $\ket{\nu_i}$ are eigenstates of bath Hamiltonian $H\sub{\bb_i}$, with energy eigenvalues $\epsilon_\mu(i)$ and $\epsilon_\nu(i)$, respectively. Such Kraus operators are constructed if, before and after the bath's joint unitary evolution with the system, we subject it to projective energy measurements.  

The full trajectory that the system takes during the protocol, therefore, can be expressed as
\be
	\Gamma = \Gamma_{(l,n_0, ..., n_N), (\mu_1, \nu_1), (\mu_2, \nu_2), ... (\mu_N, \nu_N)}, 
\ee
where $\Gamma\sub{\s} := (l,n_0, ..., n_N) \equiv \ket{\psi_l} \mapsto \ket{\tilde \psi_l} \mapsto \ket{e_{n_0}} \mapsto \dots \mapsto \ket{e_{n_N}}$ is the sequence of time-local eigenstates of the system during the protocol. Note that, here, we identify $n_0 \equiv m$ and $n_1 \equiv n$ as the eigenstate labels during Step \textbf{(III)}. The bath indices $(\mu_i, \nu_i)$ merely indicate the sequence of energy measurement outcomes on the baths, and they only contribute to the probabilities of the system trajectories $\Gamma\sub{\s}$. The probability of the trajectory $\Gamma$ is evaluated to be
\be
	P(\Gamma) && = \<\tilde \psi_l|\tilde \rho|\tilde \psi_l\> \| \mathcal{K}_\Gamma\|^2, \nonumber \\	
	 && = \<\tilde \psi_l|\tilde \rho|\tilde \psi_l\>   \, |\<\tilde \psi_l|e_{n_0}\>|^2    \prod_{i=1}^N  \<\mu_i |\tau\sub{\bb_i}| \mu_i\>  \,  |\<e_{n_i} \nu_i| \vv_i | e_{n_{i-1}} \mu_i\>|^2,  \label{eq:fullf}
\ee
where we have introduced the full Kraus operator for the protocol,
\be
	\mathcal{K}_\Gamma : = \Pi[e_{n_N}]K_{\mu_N, \nu_N} \dots \Pi[e_{n_1}]K_{\mu_1, \nu_1}\Pi[e_{n_0}] \Pi[\tilde \psi_l], \quad
\ee
with $\| \mathcal{K}_\Gamma\|:= \max_\phi \sqrt{\<\phi|\mathcal{K}_\Gamma^\dagger \mathcal{K}_\Gamma |\phi\>} = \sqrt{\<\tilde \psi_l|\mathcal{K}_\Gamma^\dagger \mathcal{K}_\Gamma|\tilde \psi_l\>}$ denoting the operator norm of $\mathcal{K}_\Gamma$. Averaging over all the measurement outcomes on the bath, meanwhile, yields the probabilities for the system-only trajectories $\Gamma\sub{\s}$, given as
\be \label{eq:coarse-grained-traj-prob}
	P(\Gamma\sub{\s}) &=& \< \tilde \psi_l| \tilde \rho |\tilde \psi_l\>  \, |\<\tilde \psi_l|e_{n_0}\>|^2 \, \prod_{i=1}^N \<e_{n_i}|\tau_i|e_{n_i}\>.
\ee

Note that we may recover the probability for any sub-trajectory of the system by summing over all other indices of \eq{eq:coarse-grained-traj-prob}. For example, summing over the indices of Steps {\bf (I)} and {\bf (IV)}, and the classical thermalization of Step {\bf (III)},  the probabilities for the system's quantum decoherence trajectories $\Gq$ are obtained as
\be
	\sum_{n_{i> 0}}     P(\Gamma\sub{\s})
	= \<\tilde \psi_l|\tilde \rho|\tilde \psi_l\>   \, |\<\tilde \psi_l|e_{n_0}\>|^2 = P\left(\Gq \right).
\ee

Summing instead over the indices of Steps {\bf (I)} and {\bf (IV)}, and the quantum decoherence of Step {\bf (III)},  the probabilities for the system's classical thermalization trajectories $\GclS$ are
\be
	\sum_{l,n_{i >1}}    P(\Gamma\sub{\s})
	&=& \sum_l \<\tilde \psi_l|\tilde \rho|\tilde \psi_l\>   \, |\<\tilde \psi_l|e_{n_0}\>|^2   \<e_{n_1}| \tau_1|e_{n_1}\>, \nonumber \\
	&=&   \<e_m| \tilde \eta |e_m\>  \<e_n| \tau_1|e_n\> = P\left(\GclS \right).
\ee

We may also reconstruct the full density operator for the system, at any point along the trajectory, see Fig.~\ref{fig:trajectories}, by weighting the pure states by the total trajectory probabilities that include this term. For example,  the average state after the decoherence process in Step {\bf (III)} is indeed
\be
	\sum_m \, |e_m \>\<e_m| \,  \sum_{l,n_1, ..., n_N}  P(\Gamma\sub{\s}) = \sum_m \, |e_m \>\<e_m| \,    \<e_m| \tilde \eta |e_m\> = \tilde \eta .
\ee

\medskip

The time-reversed trajectories can be defined by reversing the order of the protocol. Here we have Step \textbf{(IV)}: quasistatic reversed isothermal jumps $\ket{e_{n_N}} \mapsto \dots \mapsto \ket{e_{n_1}}$; Step \textbf{(III)} reversed thermalization $\ket{e_{n_1}} \mapsto \ket{e_{n_0}}$ followed by reversed decoherence $\ket{e_{n_0}} \mapsto \ket{\tilde \psi_l}$; and Step \textbf{(I)}: reversed unitary evolution $\ket{\tilde \psi_l} \mapsto \ket{\psi_l}$. Moreover, we shall consider the time-reversed thermalization maps $\Lambda_i^* : \rho \mapsto \tr\sub{\bb_i}[\vv_i^\dagger (\rho \otimes \tau\sub{\bb}) \vv_i]$. Note that the only difference between $\Lambda_i$ and $\Lambda_i^*$ is that we have applied the time reversal operation on the unitaries $\vv_i$, transforming them to $\vv_i^\dagger$.  But since the sequence of measurements on the bath during the forward protocol was $(\mu_i, \nu_i)$, we shall take the time-reversal sequence of these outcomes, namely, $(\nu_i, \mu_i)$. As such, the corresponding time-reversed Kraus operators for the thermalization channels will be 
\be
	K_{\nu_i, \mu_i}^* 
	&:=& \sqrt{\<\nu_i| \tau\sub{\bb_i}|\nu_i\>} \<\mu_i| \vv_i^\dagger |\nu_i\> \nonumber 
	=  \sqrt{\frac{\<\nu_i| \tau\sub{\bb_i}|\nu_i\>}{\<\mu_i| \tau\sub{\bb_i}|\mu_i\>}} \, K_{\mu_i, \nu_i}^\dagger = \sqrt{\frac{q_{n_{i-1}}^{(i)}}{q_{n_{i}}^{(i)}}} \, \, K_{\mu_i, \nu_i}^\dagger ,
\ee
where $q_{n_{i}}^{(j)} := \<e_{n_i}| \tau_j | e_{n_i}\>$. Here we have used the fact that, given the energy conservation of the thermalization unitary $\vv_i$, it follows that 
\be
	\frac{\<\nu_i| \tau\sub{\bb_i}|\nu_i\>}{\<\mu_i| \tau\sub{\bb_i}|\mu_i\>} 
	&=& e^{(\epsilon_\mu(i) - \epsilon_\nu(i))/k_B T} = e^{(E_{n_i}^{(i)} - E_{n_{i-1}}^{(i)})/k_B T}= \frac{\<e_{n_{i-1}}|\tau_{i}|e_{n_{i-1}}\>}{\<e_{n_{i}}|\tau_{i}|e_{n_{i}}\>} 
	= \frac{q_{n_{i-1}}^{(i)}}{q_{n_{i}}^{(i)}},
\ee
where $E_{n_i}^{(j)} := \<e_{n_i}| H^{(j)} |e_{n_i}\>$.   Finally, the time-reversed trajectories can be denoted as 
\be
	\Gamma^* = \Gamma^*_{(n_N,\dots, n_0, l), (\nu_1, \mu_1), (\nu_2,\mu_2),\dots, (\nu_N, \mu_N) }, 
\ee
which occur with the probability 
\be  \label{eq:backprobs}
	P^*(\Gamma^*) &=& \<e_{n_N}| \tau_N| e_{n_N}\> \| \mathcal{K}_{\Gamma^*}\|^2 ,  \label{eq:fullb}
\ee
where we introduce the time reversed Kraus operators for the full protocol, 
\be \label{eq:backK}
\mathcal{K}_{\Gamma^*} : = \sqrt{\frac{\prod_{i=1}^N q_{n_{i-1}}^{(i)}}{\prod_{i=1}^N q_{n_{i}}^{(i)}}} \, \,  \mathcal{K}_\Gamma^\dagger.
\ee

Now we may evaluate the entropy production for the full protcol, which is given by  \eq{eq:fullf} and \eq{eq:backprobs} to be
\begin{align}
	s_\mathrm{irr}(\Gamma)
	&:= k_B \,  \log{\frac{P(\Gamma)}{P^*(\Gamma^*)}}, \nonumber \\
	& =  k_B \, \log{\frac{\<\tilde \psi_l|\tilde \rho|\tilde \psi_l\> \| \mathcal{K}_\Gamma\|^2}{\<e_{n_N}|\tau_N|e_{n_N}\>  \| \mathcal{K}_{\Gamma^*}\|^2}} = k_B \log{\frac{\<\tilde \psi_l|\tilde \rho|\tilde \psi_l\>}{\<e_{n_N}|\tau_N|e_{n_N}\>}} + k_B \sum_{i=1}^N \log{\frac{q_{n_i}^{(i)}}{q_{n_{i-1}}^{(i)}}}, 
\end{align}
where we have used the fact that $\| \mathcal{K}_\Gamma\|^2 = \|\mathcal{K}_\Gamma^\dagger\|^2$. Note that the entropy production is independent of the bath measurement results. In other words, the entropy production can be purely determined by the system trajectories $\Gamma\sub{\s}$. 

It is trivial to show that this entropy production can be split into the three terms
\be
	s_\mathrm{irr}(\Gamma)
	&=& \sqG + \sclG + s_\mathrm{irr}^\mathrm{cl}\left(\Gamma^{\bf{(IV)}}\right), \quad \quad \quad
\ee
where  $\sqG$ and $\sclG$ are defined in \eq{eq:quantum-entropy-production} and \eq{eq:classical-entropy-production}, respectively, and 
\be\label{eq:ent-prod-IV}
	s_\mathrm{irr}^\mathrm{cl}\left(\Gamma^{\bf{(IV)}}\right) 
	&:=& k_B \, \log{\frac{q_{n_1}^{(1)}}{q_{n_N}^{(N)}}}  + k_B \sum_{i=2}^N \log{\frac{q_{n_i}^{(i)}}{q_{n_{i-1}}^{(i)}}} =  \sum_{i=2}^N k_B \log{\frac{q_{n_{i-1}}^{(i-1)}}{q_{n_{i-1}}^{(i)}}} 
\ee
is the entropy production of Step \textbf{(IV)}.

Since the average entropy production is additive, i.e $\avg{s_\mathrm{irr}}_\Gamma = \qent + \clent + \avg{s_\mathrm{irr}^\mathrm{cl}}_{\Gamma^{\bf{(IV)}}}$, we will compute each term separately. Let us first turn to the last term, namely, the entropy production in Step \textbf{(IV)}. We verify that averaging over the trajectory probabilities, one obtains
\be
	\frac{\avg{s_\mathrm{irr}^\mathrm{cl}}_{\Gamma^{\bf{(IV)}}}}{k_B} 
	&=& \sum_{i=2}^N \sum_{n_{i-1}} q_{n_{i-1}}^{(i-1)} \log{\frac{q_{n_{i-1}}^{(i-1)}}{q_{n_{i-1}}^{(i)}}}   = \sum_{i=2}^N D[\tau_{i-1} \| \tau_i].
\ee
When Step \textbf{(IV)} approaches the quasistatic limit, we will have $\sum_{i=2}^N D[\tau_{i-1}\| \tau_i] \to 0$, and so $\avg{s_\mathrm{irr}}_\Gamma = \qent + \clent$.  

Now we turn to the average entropy production during Step \textbf{(III)}. Using \eq{eq:qtraj-probability} and \eq{eq:quantum-entropy-production},   and introducing the labels $p_l := \langle \tilde \psi_l | \tilde \rho |\tilde \psi_l \rangle$ and $r_m := \langle e_m |\tilde \eta | e_m \rangle$, the average quantum entropy production can be shown to be
\be  
	\qent 
	&=& \sum_{l, m} P\left(\Gq\right) \, \sqG, \nonumber \\
	&=& k_B  \sum_{l, m} p_l \,  |\langle e_m |\tilde \psi_l \rangle|^2 \,  \log {p_l   \over r_m} = k_B \, D[\tilde \rho \| \tilde \eta], \quad \quad \quad
\ee 
as stated in the main text. Here, we used the fact that $\sum_{l} p_l |\langle e_m |\tilde \psi_l \rangle|^2 \log{r_m} = r_m \log{r_m}$, and that $\tr[\tilde \eta \log{\tilde \eta}] = \tr[\tilde \rho \log{\tilde \eta}]$. 
Meanwhile, the average classical entropy production is given by \eq{eq:cltraj-probability} and \eq{eq:classical-entropy-production} as
\be  
	\clent 
	&=& \sum_{m, n} P\left(\GclS\right) \, \sclG 
	= k_B  \sum_m r_m \log{r_m \over q_m}
	= k_B D[\tilde \eta \| \tau_1],
\ee
where here $q_m := \<e_m| \tau_1| e_m\>$.

\subsection*{Fluctuations in quantum and classical heat}\label{app:variance-quantum-heat}

Here, we shall provide expressions for the fluctuations in quantum and classical heat during the thermalization process in Step {\bf (III)} of the work extraction protocol. For notational simplicity, we shall denote the Hamiltonian as  $H = \sum_{m=1}^d E_m \pr{e_m}$, the initial state of the system as $\rho = \sum_{l=1}^d p_l \pr{\psi_l}$, its state after decoherence as   $\eta := \sum_m r_m \pr{e_m}$, and its thermal state as $\tau := \sum_n q_n \pr{e_n}$.

As the system decoheres with respect to the Hamiltonian, we obtain trajectories $\Gamma^\mathrm{q}_{(l,m)}:= \ket{\psi_l} \mapsto \ket{e_m}$, with probabilities $P(\Gamma^\mathrm{q}_{(l,m)}) = p_l |\<\psi_l|e_m\>|^2$ and quantum heat $\Qq(\Gamma^\mathrm{q}_{(l,m)}):= \<e_m| H | e_m\> - \<\psi_l|H | \psi_l\>$. The average quantum heat for a decoherence process is always zero, 
\be
 	\avg{\Qq}
	&:=& \sum_{l,m} P(\Gamma^\mathrm{q}_{(l,m)}) \, \Qq(\Gamma^\mathrm{q}_{(l,m)}) 
	= \sum_{m} \<e_m| \rho | e_m\> \, \<e_m| H | e_m\>  - \tr[H \, \rho]= 0.
\ee
Hence the variance in quantum heat is equal to its second moment:
\be
	\Var{\Qq} &:=& \avg{\Qq^2} - \avg{\Qq}^2 = \avg{\Qq^2}  := \sum_{l,m} P(\Gamma^\mathrm{q}_{(l,m)}) \, \Qq^2(\Gamma^\mathrm{q}_{(l,m)}), \nonumber \\
	&=& \sum_{l,m} p_l |\<\psi_l|e_m\>|^2 \<e_m| H^2 | e_m\> + \sum_l p_l \<\psi_l|H | \psi_l\>^2   - 2\sum_{l,m} p_l |\<\psi_l|e_m\>|^2 \<\psi_l|H | \psi_l\> \<e_m| H | e_m\>. 
\ee
Noting that $\sum_m |\<\psi_l|e_m\>|^2 \<e_m| H^k | e_m\> = \<\psi_l| H^k |\psi_l\>$, the variance in quantum heat reduces to 
\be \label{eq:average-variance-Hamiltonian}
	 \Var{\Qq} &=& \sum_l p_l \left( \<\psi_l|H^2|\psi_l\> - \<\psi_l|H|\psi_l\>^2 \right)= \sum_l p_l \, \Delta(H, \psi_l),
\ee
where $\Delta(H, \rho) := \tr[H^2 \rho] - \tr[H \rho]^2$ is the variance of the Hamiltonian $H$ in state $\rho$. In other words, the variance in quantum heat is the average variance of the Hamiltonian in the pure state components of the initial state $\rho$.

We now give upper and lower bounds to the variance in quantum heat.  For the upper bound we have 
\be  \label{eq:quantum-heat-variance-uper-bound}
	\Delta(H, \rho) &-& \Var{\Qq} 
	= \sum_l p_l \<\psi_l|H|\psi_l\>^2 - \tr[H \rho]^2   = \sum_l p_l (\<\psi_l| H|\psi_l\> - \tr[H \rho])^2 \geqslant 0. 
\ee
To obtain a lower bound, we use the fact that $\Delta(H, \rho) = I_\alpha(H, \rho)$ whenever $\rho$ is a pure state,  where $I_\alpha(H, \rho) = \tr[H^2 \rho] - \tr[H \rho^\alpha H \rho^{1-\alpha} ]$ for $\alpha \in (0,1)$ is the Wigner-Yanase-Dyson skew information of the observable $H$ in $\rho$ \cite{Wigner1963}. Using the Lieb concavity theorem \cite{Lieb1973} it follows that
\be \label{eq:quantum-heat-variance-lower-bound}
	\Var{\Qq} &=& \sum_l p_l \, I_\alpha(H, \psi_l) \geqslant I_\alpha(H, \rho).
\ee
Combining \eq{eq:quantum-heat-variance-uper-bound} and \eq{eq:quantum-heat-variance-lower-bound} shows that the variance in quantum heat obeys
\be \label{eq:appendix-qheat-var-inequality}
	\Delta(H, \rho) \geqslant  \Var{\Qq} \geqslant I_\alpha(H, \rho),
\ee
where the equalities are saturated  if $\rho$ is pure.

As the system thermalizes, we obtain trajectories $\Gamma^\mathrm{cl}_{(m,n)}:= \ket{e_m} \mapsto \ket{e_n}$, with probabilities $P(\Gamma^\mathrm{cl}_{(m,n)}) = r_m q_n$ and classical heat $\Qcl(\Gamma^\mathrm{cl}_{(m,n)}):= \<e_n| H | e_n\> - \<e_m|H | e_m\>$. The average classical heat is therefore
\begin{align}
\avg{\Qcl} &:= \sum_{m,n} P(\Gamma^\mathrm{cl}_{(m,n)}) \Qcl(\Gamma^\mathrm{cl}_{(m,n)}) = \sum_{m,n} r_m q_n \tr[H(\pr{e_n} - \pr{e_m})]= \tr[H(\tau - \eta)] \equiv \tr[H(\tau - \rho)],
\end{align}
while the second moment is 
\begin{align}
\avg{\Qcl^2} := \sum_{m,n} P(\Gamma^\mathrm{cl}_{(m,n)}) \Qcl(\Gamma^\mathrm{cl}_{(m,n)})^2 &= \sum_{m,n} r_m q_n \bigg(\tr[H \pr{e_n}]^2 + \tr[H \pr{e_m}]^2 - 2\tr[H \pr{e_n}]\tr[H \pr{e_m}] \bigg), \nonumber \\
&= \tr[H^2(\tau + \eta)] - 2\tr[H \tau]\tr[H \eta] \equiv \tr[H^2(\tau + \rho)] - 2\tr[H \tau]\tr[H \rho].
\end{align}
Note that here we have used the fact that $\tr[H^k \eta] = \sum_n \tr[H^k \pr{e_n} \rho \pr{e_n}] = \tr[H^k \rho] $.

The variance in classical heat, therefore, is 
\begin{align}\label{eq:app-var-classical-heat}
\Var{\Qcl}:= \avg{\Qcl^2} - \avg{\Qcl}^2 = \Delta(H, \eta) + \Delta(H, \tau)\equiv \Delta(H, \rho) + \Delta(H, \tau).
\end{align}

\subsection*{Quantum and classical heat variances for a qubit} \label{sec:qubitheat}
Let us first consider the variance in quantum heat for the decoherence trajectories  $\Gqwoindex$ of  a qubit in state $\rho_\tth = p \Pi[\tth_-] + (1-p) \Pi[\tth_+]$ and with Hamiltonian $H^{(1)} = \frac{\hbar \omega_1}{2}(\Pi[e_+] - \Pi[e_-])$. One finds that when $d=2$, the matrix elements of the doubly stochastic matrix $\bm{M}(\Theta)$ are $M_{k\ne l,l}^{(\Theta)} = \frac{1}{2}\sin^2(\Theta \pi/2)$, and $M_{l,l}^{(\Theta)} = 1 - \frac{1}{2}\sin^2(\Theta \pi/2)$. By solving the equation $|\tth| = 2 \sin^{-1} \big(\sin(\Theta \pi/2)/\sqrt{2}\big)$, we may equivalently write these as  $M_{k\ne l,l}^{(\tth)} = \sin^2(\tth/2) \equiv \coh$, and $M_{l,l}^{(\tth)} = 1 - \sin^2(\tth/2) \equiv 1 - \coh$, as defined in \eq{eq:qubitcoh}. We may therefore rewrite \eq{eq:variance-Theta} as
\begin{align}\label{eq:coh-variance-qubit}
\Delta(H^{(1)}, \tth_\pm) &= \coh \sum_{k \in \pm} \bigg( E_k^{(1)} -  E^{(1)}_\pm \bigg)^2 - \coh^2  \bigg( \sum_{k\in \pm} \big(E_k^{(1)} - E^{(1)}_\pm \big) \bigg)^2, \nonumber \\
& = \left(\hbar \omega_1 \right)^2 \left(\coh - \coh^2\right) \equiv \frac{\left(\hbar \omega_1 \right)^2}{4} \sin^2(\tth).
\end{align}
In the second line, we have used the fact that for the qubit model,  $E_k^{(1)} - E^{(1)}_\pm \in \{0, \pm \hbar \omega_1\}$. Since the variance of the Hamiltonian is the same for both eigenstates of the qubit, it follows that the variance in quantum heat is always 
\be
\Var{\Qq} = \Delta(H^{(1)}, \tth_\pm)  =  \frac{\left(\hbar \omega_1 \right)^2}{4} \sin^2(\tth),
\ee
which monotonically increases as $|\tth|$ increases from $0$ to $\pi/2$.

Let us now consider the variance in classical heat  for the  thermalization trajectories $\Gclwoindex$. Note that there are only two trajectories which contribute non-vanishing values of classical heat: $\ket{e_-} \mapsto \ket{e_+}$, with absorbed heat $\hbar \omega_1$, occurring with probability $r_\tth(1-q_1)$ with $q_1\ge 1/2$; and $\ket{e_+} \mapsto \ket{e_-}$, with absorbed heat $- \hbar\omega_1$, occurring with probability $(1- r_\tth)q_1$. From \eq{eq:app-var-classical-heat}, we can obtain the simplified expression for the classical heat variance as
\be
	\Var{\Qcl} 
	&=&  \Delta(H^{(1)}, \eta_\tth) + \Delta(H^{(1)}, \tau_1), \nonumber \\
	&=& (\hbar \omega_1)^2 \, ( r_\tth - r_\tth^2 ) +  (\hbar \omega_1)^2 (q_1 - q_1^2), 
\ee
where $r_\tth = q_1 \, \exp(-\noneq(\rho_\tth)) \geqslant 1/2$ is a function of the non-thermality of the state $\rho_\tth$. Hence $\Var{\Qcl}$ monotonously increases with $\noneq(\rho_\tth)$, see also Fig.~\ref{fig:CQfluc}.

\subsection*{Hamiltonian-covariant channels and energy coherence for qubits}\label{sec:dephasing-coherence-qubits}

In order to see how Hamiltonian-covariant channels affect the energy coherence of the eigenbasis of $\rho$, it will be useful to work in the geometric picture of the Bloch sphere, where $\rho = \frac{1}{2}(\one + \vec n. \vec \sigma)$ and $H = \frac{\hbar \omega}{2} \sigma_3$. Here $\vec n := (n_1,n_2,n_3)$ is the Bloch vector such that $n_i \in \re $ and  $|\vec n| \leqslant 1$, and $\vec \sigma := (\sigma_1,\sigma_2,\sigma_3)$ with $\sigma_i$ the Pauli matrices. As such, the spectral projections of $H$ and $\rho$  can be expressed as 
\begin{align}
&\pr{e_\pm}:= \frac{1}{2}\left(\one \pm \sigma_3\right), &\pr{\theta_\pm}:= \frac{1}{2}\left(\one \pm \frac{\vec n}{|\vec n|}. \vec \sigma \right),
\end{align}
which give the energy coherence of the eigenbasis of $\rho$ as
\begin{align}
\coh(\rho) := \min_{i,j} \tr[\pr{e_i}\pr{\theta_k}]= \frac{1}{2}\left(1 - \frac{|n_3|}{|\vec n|}\right). 
\end{align}
In other words, the energy coherence decreases as the fraction of the Bloch vector along the Hamiltonian axis $x_3 := (0,0,1)$ increases, where we note that here, we define $0/0 := \lim_{x\to 0}  x/x = 1$, meaning that the energy coherence of the complete mixture is zero. Now let us consider the two states $\rho = \frac{1}{2}(\one + \vec n. \vec \sigma)$ and $\e(\rho) = \frac{1}{2}(\one + \vec m. \vec \sigma)$. We therefore have  
 \begin{align}
\coh(\rho) - \coh(\e(\rho)) = \frac{1}{2}\left(\frac{| m_3|}{|\vec{m}|} - \frac{ |n_3|}{|\vec{n}|} \right)  \geqslant 0 \iff \frac{| m_3|^2}{|\vec{m}|^2} \geqslant \frac{ |n_3|^2}{|\vec{n}|^2}.
\end{align}
Now we wish to see what subset of Hamiltonian-covariant channels $\e$ will guarantee that $\coh(\rho) - \coh(\e(\rho)) \geqslant 0$ for all $\rho$.

Due to the convex structure of quantum channels \cite{Heinosaari2011}, any quantum channel that maps from a $d$-dimensional Hilbert space to itself can be constructed as a convex combination of ``extremal'' quantum channels $ \{\e_i \}$  where extremality of $\e_i$ is defined as $\e_i = \lambda \e_j + (1-\lambda) \e_k$, with $\lambda \in [0,1]$, only if $\e_j = \e_k = \e_i$. In the special case of $d=2$, as shown in Corollary 15 of Ref. \cite{Friedland2016}, a quantum channel $\e_i$ is extremal if either $\e_i$ is unitary, or it is not a convex combination of unitary channels and the rank of its corresponding Choi-state is 2. The Choi-state associated with a qubit quantum channel $\e$ is defined as 
\be\label{eq:Choi-state}
\varrho\sub{\e} := (\e \otimes \one)\pr{\Phi_+},
\ee
where $\ket{\Phi_+} := \frac{1}{\sqrt{2}}(\ket{\varphi_+,\varphi_+} + \ket{\varphi_-,\varphi_-} )$ with $\{\ket{\varphi_\pm}\}$ any orthonormal basis of $\co^2$. Therefore, we may always write a qubit  channel $\e$ as 
\begin{align}\label{eq:decomposition-qubit-channel}
\e(\rho) &= \lambda\,  \uu(\rho) + (1-\lambda) \T(\rho),
\end{align}
where $\lambda \in [0,1]$ and
\begin{align}
\uu(\rho) & = \sum_{j} p_j \, \uu_j( \rho),  &\T(\rho) = \sum_{k} q_k \, \T_k(\rho),
\end{align}
with $p_j, q_k >0$ and $\sum_{j} p_j =\sum_{k} q_k = 1$. Moreover,  $\uu_j(\rho) := U_j\rho U_j$ with $U_j$ unitary operators, and $\T_k(\rho) = \sum_{l\in \pm} K_{l,k} \rho K_{l,k}^\dagger$, with Kraus operators $ K_{+,k}= |\psi_k\>\<\varphi_+| , K_{-,k} = |\psi_k'\>\<\varphi_-|$, where $\{\ket{\psi_k}, \ket{\psi_k'}\}$ are any pair of pure states, not necessarily orthogonal. It is simple to verify, by \eq{eq:Choi-state} and the definition of the Kraus operators above, that $\T_k$ have the Choi states $\varrho\sub{\T_k} = \frac{1}{2} (\pr{\psi_k,\varphi_+} +\pr{\psi_k',\varphi_-} )$, which are rank-2 and thus satisfy the extremality condition.

Now let us assume that $\e$ is covariant with respect to the Hamiltonian $H$, i.e. for any $\rho$  and $t \in \re$, we have $e^{-i t H}\e(\rho)e^{i t H} = \e(e^{-i t H} \rho e^{i t H})$. Of course, this means that $\uu_j$ and $\T_k$ are also Hamiltonian-covariant, implying that $U_j = e^{ i \phi_j \sigma_3}$, so that $\uu$ is a probabilistic rotation about the Hamiltonian axis $x_3$. As for  $\T$, let us note that 
\begin{align}
&e^{-i t H}\T_k(\rho)e^{i t H} = \T_k(e^{-i t H} \rho e^{i t H}) \implies \nonumber \\
&e^{-i t H}|\psi_k\>\<\psi_k|e^{i t H} \<\varphi_+|\rho|\varphi_+\> + e^{-i t H}|\psi_k'\>\<\psi_k'|e^{i t H} \<\varphi_-|\rho|\varphi_-\> \nonumber \\
& \qquad \qquad = |\psi_k\>\<\psi_k|  \<\varphi_+|e^{-i t H}\rho e^{i t H}|\varphi_+\> + |\psi_k'\>\<\psi_k'| \<\varphi_-|e^{-i t H}\rho e^{i t H} |\varphi_-\>,
\end{align}
 implies that $\{\ket{\varphi_\pm}\} \equiv \{\ket{e_\pm}\}$, while $\ket{\psi_k}, \ket{\psi_k'}$ must also be eigenstates of $H$ although, as stated before, they may be the same eigenstate. Therefore, there are only three extremal channels $\T_k$: $\T_1(\rho) = \frac{1}{2}(\one + \sigma_3)$, $\T_2(\rho) = \frac{1}{2}(\one - \sigma_3)$, and $\T_3(\rho) = \frac{1}{2}(\one -n_3 \sigma_3)$. Consequently, $\T(\rho) =  \frac{1}{2}(\one + v\sigma_3)$, with $v = q_1 - q_2 - q_3 n_3$.

 It trivially follows that 
\begin{align}\label{}
|m_3|^2 &= |\tr[\sigma_3 \e(\rho)]|^2 = |\lambda \tr[\sigma_3 \uu(\rho)] + (1-\lambda) \tr[\sigma_3 \T(\rho)]|^2 = |\lambda n_3 + (1-\lambda) v|^2 = \lambda^2  \beta^2 |n_3|^2,
\end{align}
where 
\begin{align}
\beta = \left(1 + \frac{1 - \lambda}{\lambda}\frac{v}{n_3}\right).
\end{align}
Moreover, denoting $m_\perp = (m_1,m_2,0)$ as the component of $\vec m$ that is orthogonal to $x_3$, so that $|\vec m|^2 = |m_3|^2 + |m_\perp|^2$, and similarly with $n_\perp$, we obtain 
\begin{align}
|m_\perp|^2 &=   |\tr[(\sigma_1 + i\sigma_2) \e(\rho)]|^2 = \lambda^2 |\tr[(\sigma_1 + i\sigma_2) \uu(\rho)]|^2 =\delta \lambda^2 |\tr[(\sigma_1 + i\sigma_2) \rho]|^2 = \delta \lambda^2 |n_\perp|^2, 
\end{align}
where $\delta \in [0,1]$, with $\delta = 1$ if $\uu(\rho) =e^{i \phi \sigma_3} \rho e^{-i \phi \sigma_3}$, and $\delta = 0$ when $\uu(\rho) = \int_{[0,2\pi)}  d\mu(\phi) e^{i \phi \sigma_3} \rho ^{-i \phi \sigma_3} \equiv \sum_{k \in \pm} \pr{e_k} \rho \pr{e_k}$ with $\mu$ the Haar measure over $[0, 2\pi)$.

As such, we may write 
\begin{align}
\frac{|m_3|^2}{|\vec m|^2} = \frac{|m_3|^2}{|m_3|^2 + |m_\perp|^2} =  \frac{|m_3|^2}{|m_3|^2 + \delta \lambda^2 |n_\perp|^2}  = \frac{\beta^2 |n_3|^2 }{\beta^2 |n_3|^2  + \delta |n_\perp|^2}.
\end{align}
Consequently, so long as $ \beta^2 \geqslant \delta$, we have 
\begin{align}
\frac{|m_3|^2}{|\vec m|^2} \geqslant \frac{ |n_3|^2 }{|\vec n_3|^2} \implies \coh(\rho) \geqslant \coh(\e(\rho)).
\end{align}

A sufficient condition to ensure that $\coh(\rho) \geqslant \coh(\e(\rho))$ for all $\rho$, irrespective of the value of $\lambda$ and $\delta$, is if $\T$ is a depolarizing channel, i.e. $\T(\rho) = \frac{1}{2}\one$ for all $\rho$. In this case, $v=0$ and so $\beta^2 =1 \geqslant \delta$.

\newpage

%
%
%
%
%
%

{\bf Acknowledgments.}
We have the pleasure to thank Karen Hovhannisyan, Harry Miller, Cyril Elouard, Ian Ford and Bruno Mera for inspiring discussions. This research was  supported in part by the COST network MP1209 ``Thermodynamics in the quantum regime" and by the National Science Foundation under Grant No. NSF PHY-1748958.  M.H.M. acknowledges support from EPSRC via Grant No. EP/P030815/1, as well as the Slovak Academy of Sciences   under MoRePro project OPEQ (19MRP0027). A.A. acknowledges the Agence Nationale de la Recherche under the Research Collaborative Project ``Qu-DICE" (ANR-PRC-CES47). J.A. acknowledges support from EPSRC (grant EP/R045577/1) and the Royal Society. 


\bibliography{References}

\end{document}